\documentclass[aps,prd,twocolumn,showpacs]{revtex4}
\usepackage[hyperindex,breaklinks]{hyperref}
\usepackage{graphicx}% Include figure files
\def\figcoD{%
\centerline{{\bf(a)}\hskip-15pt\scalebox{0.5}{\includegraphics{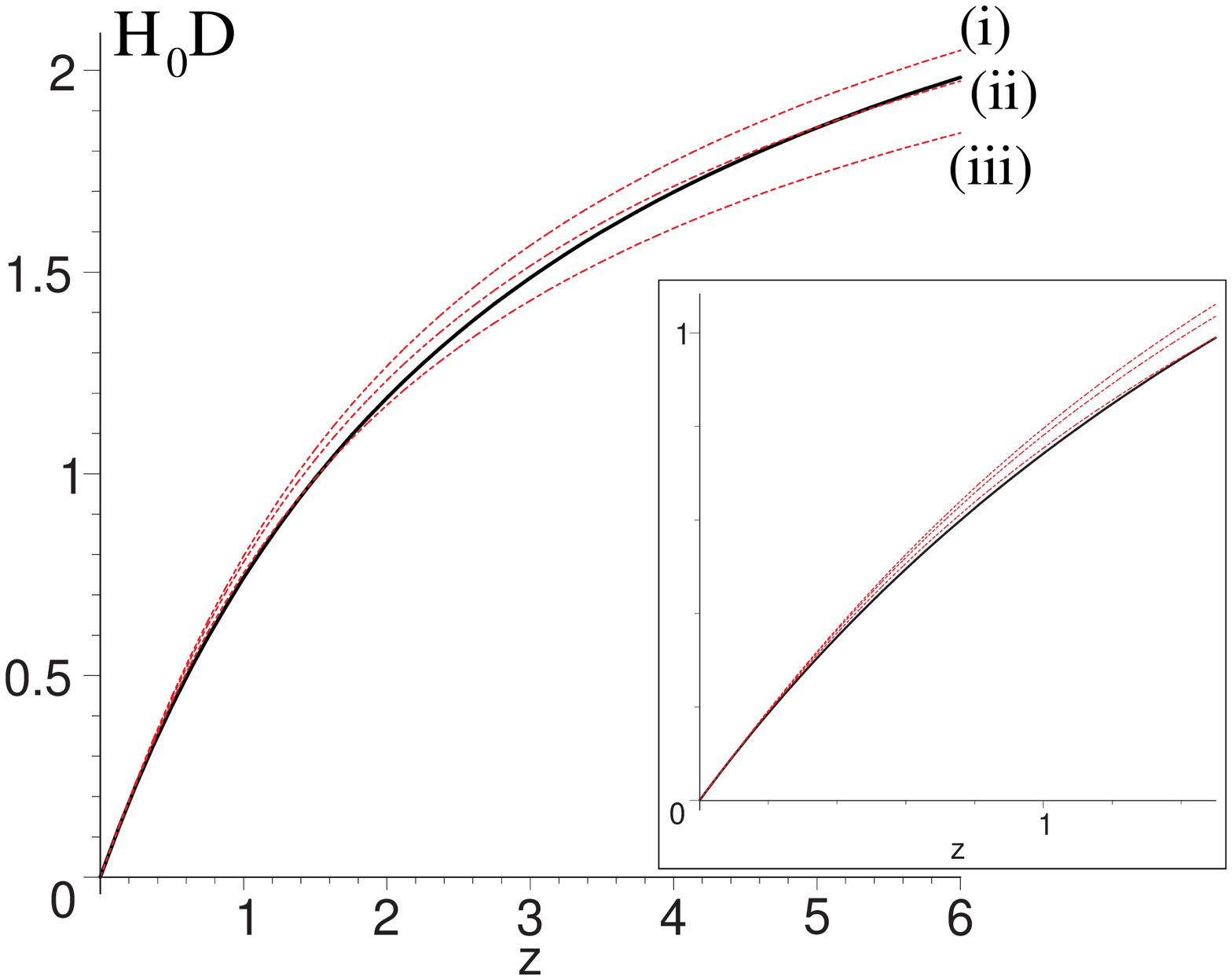}}
\quad{\bf(b)}\hskip-15pt\scalebox{0.5}{\includegraphics{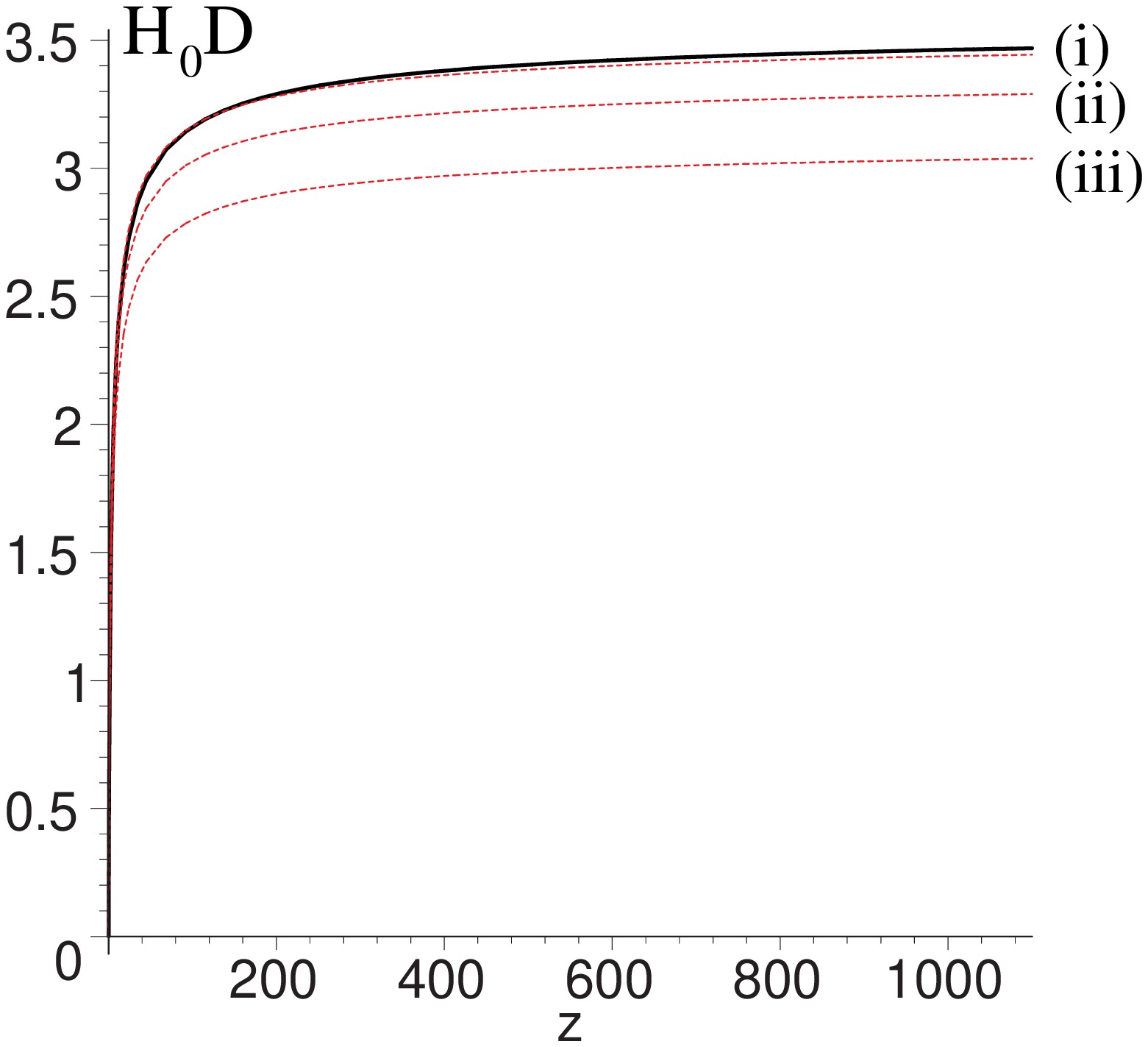}}}}
\def\figomq{\centerline{\scalebox{0.75}{\includegraphics{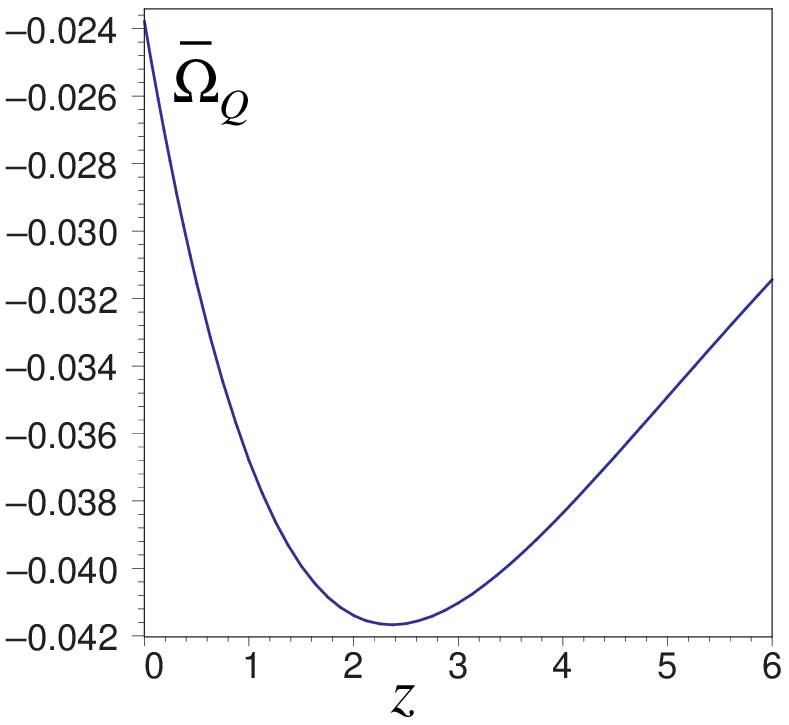}}}}
\def\figwza{\centerline{\scalebox{0.75}{\includegraphics{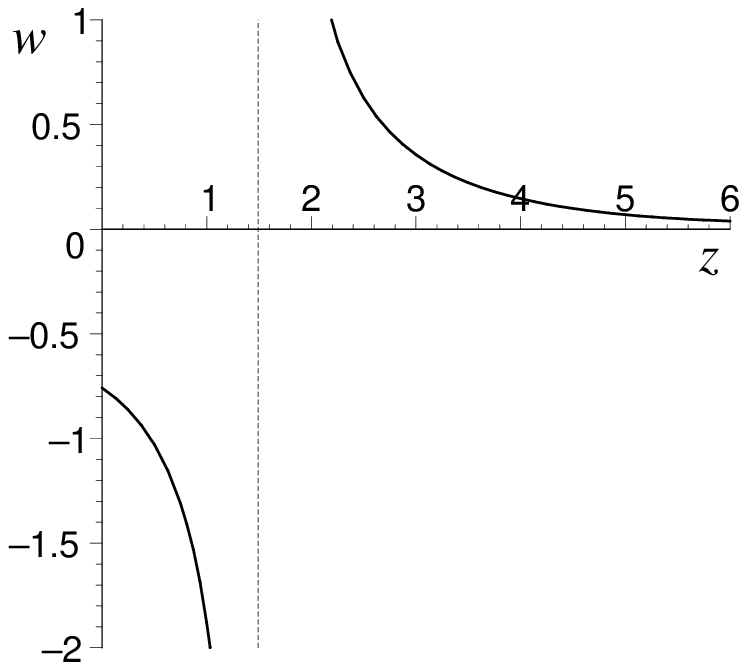}}}
\vskip-20pt\leftline{\bf(a)}\vskip10pt}
\def\figwzb{\centerline{\scalebox{0.75}{\includegraphics{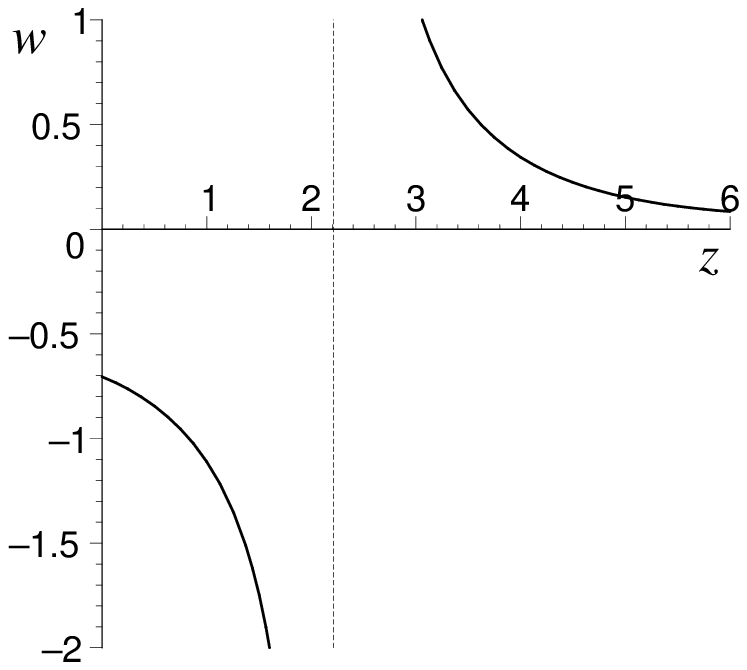}}}
\vskip-20pt\leftline{\bf(b)}\vskip5pt}
\def\figwz{\centerline{\scalebox{0.75}{\includegraphics{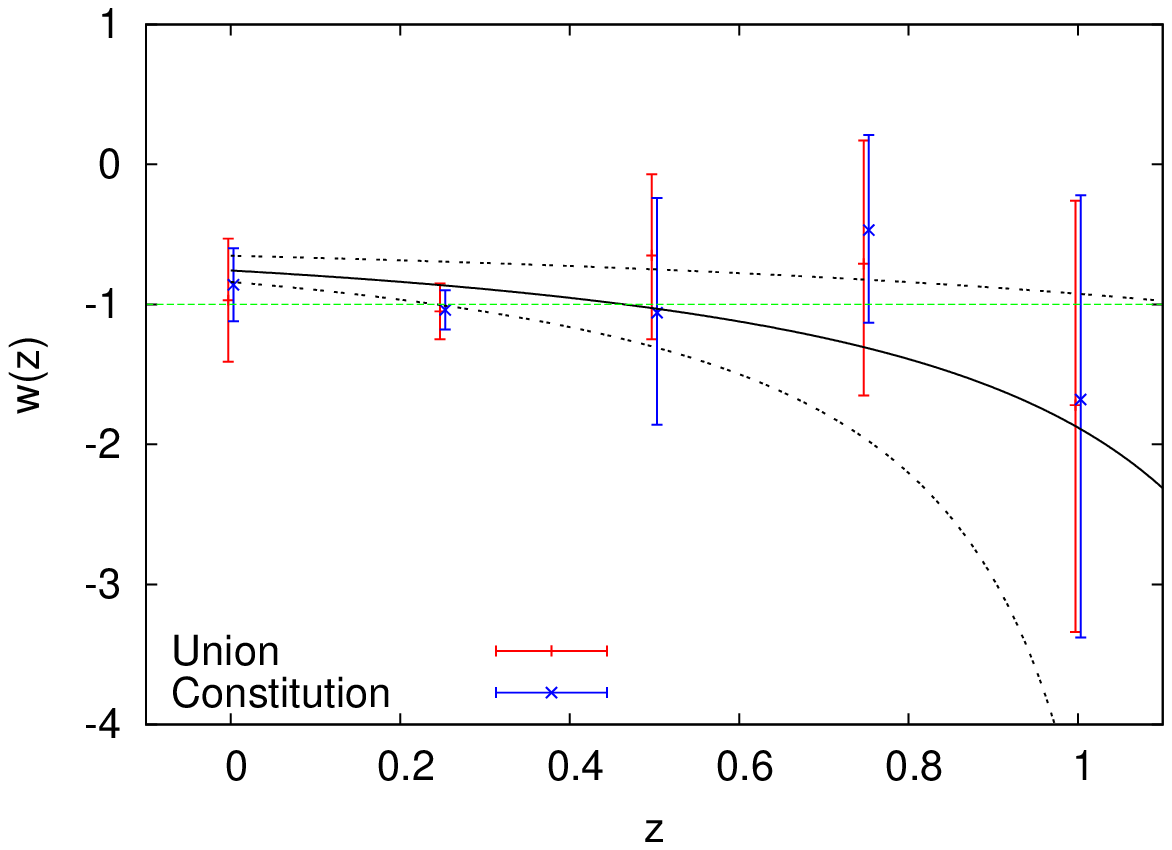}}}}
\def\figBex{\centerline{\scalebox{0.75}{\includegraphics{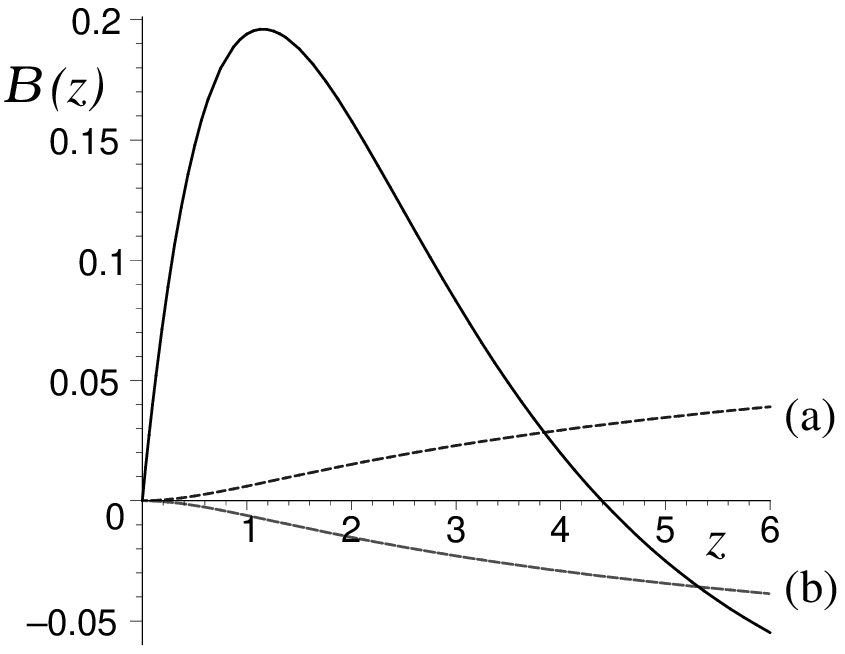}}}}
\def\figCex{\centerline{\scalebox{0.75}{\includegraphics{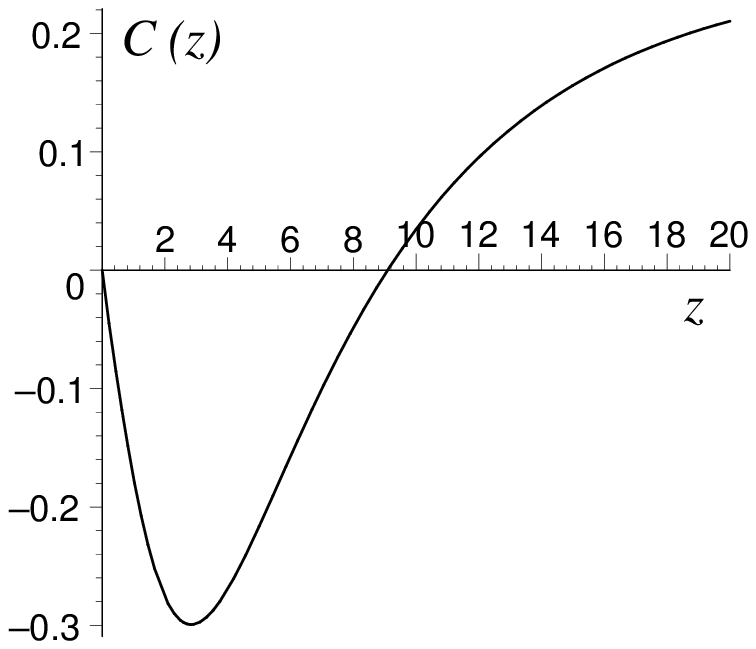}}}}
\def\figzdot{\centerline{\scalebox{0.75}{\includegraphics{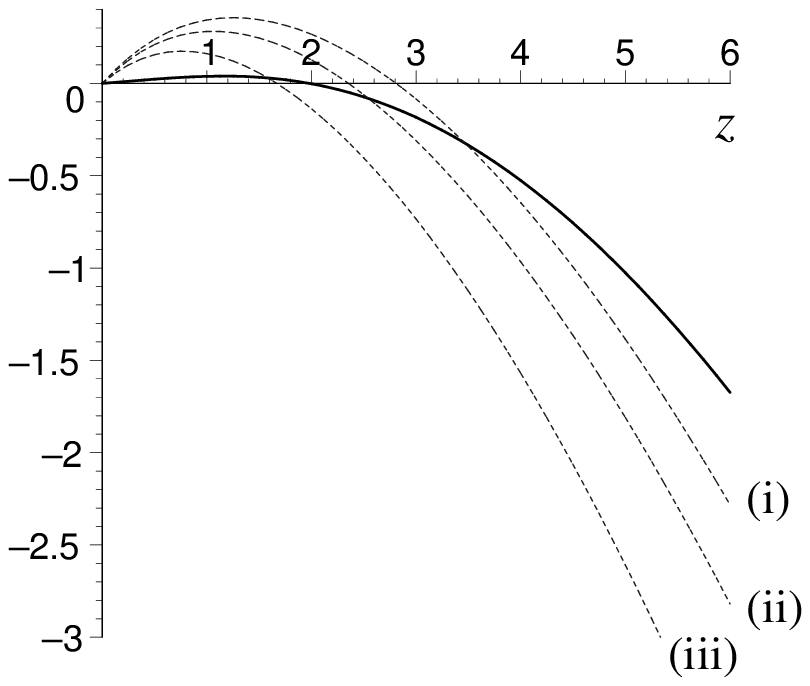}}}}
\def\figAP{\centerline{{\bf(a)}\hskip-15pt
\scalebox{0.75}{\includegraphics{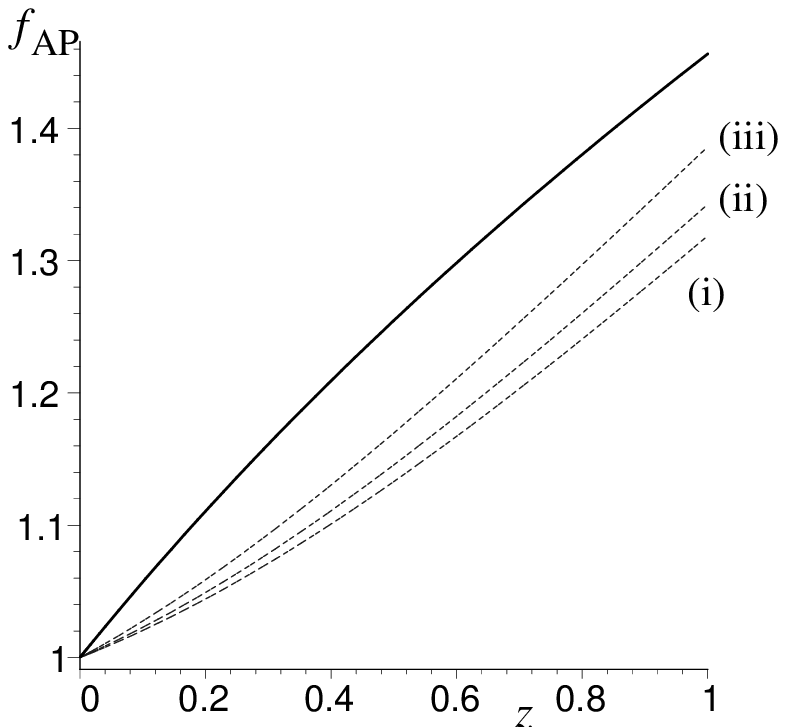}}}\smallskip
\centerline{{\bf(b)}\hskip-15pt
\scalebox{0.75}{\includegraphics{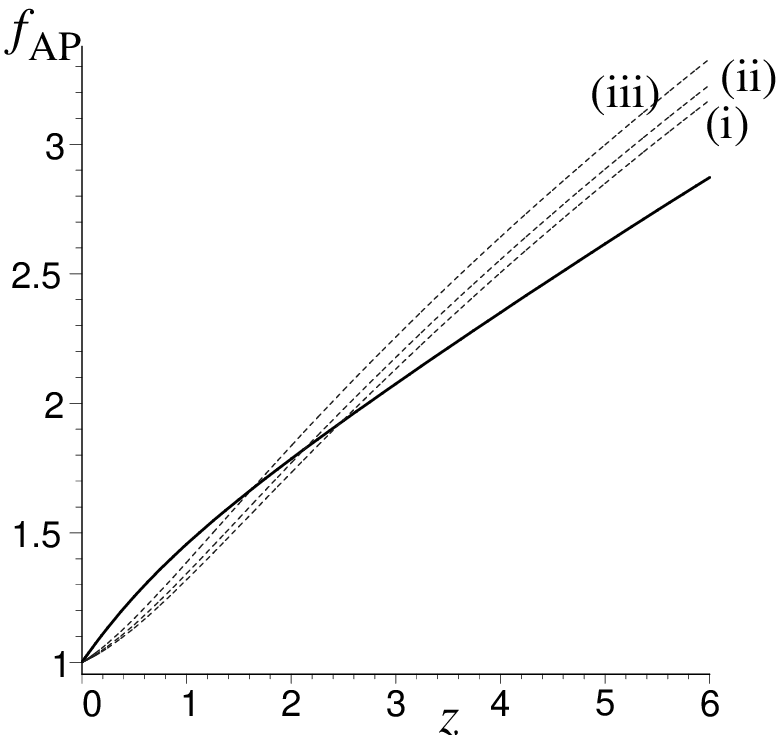}}}}
\def\figangd{\centerline{\scalebox{0.75}{\includegraphics{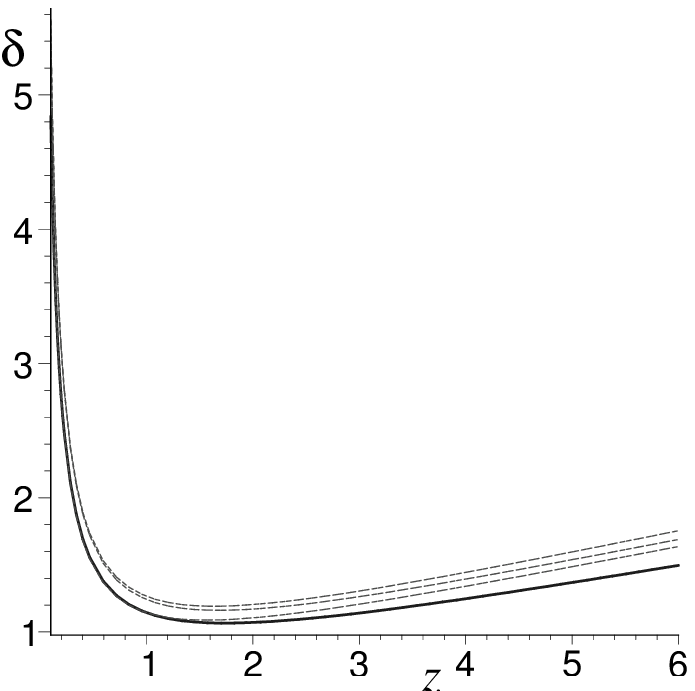}}}}
\def\figHH0{\centerline{\scalebox{0.75}{\includegraphics{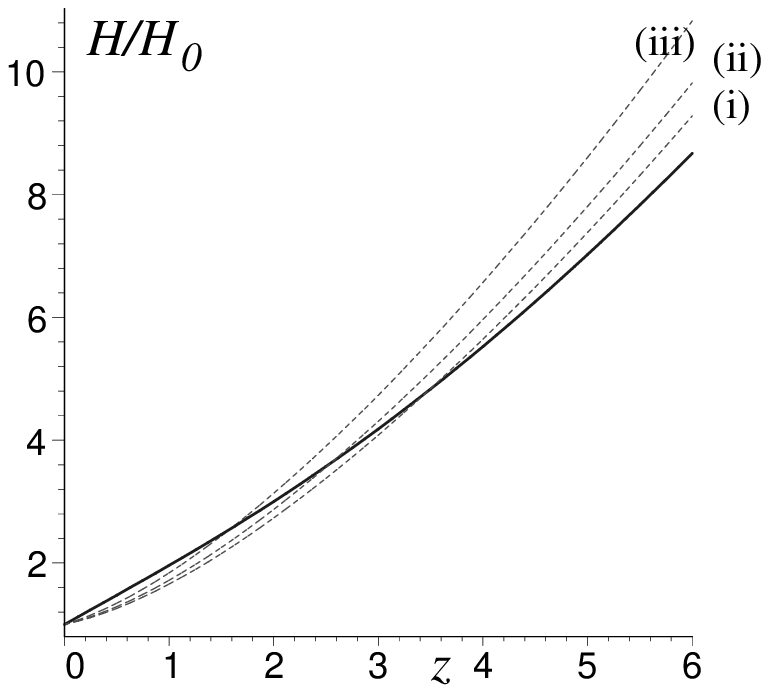}}}}
\def\figOm{\centerline{{\bf(a)}\hskip-15pt
\scalebox{0.65}{\includegraphics{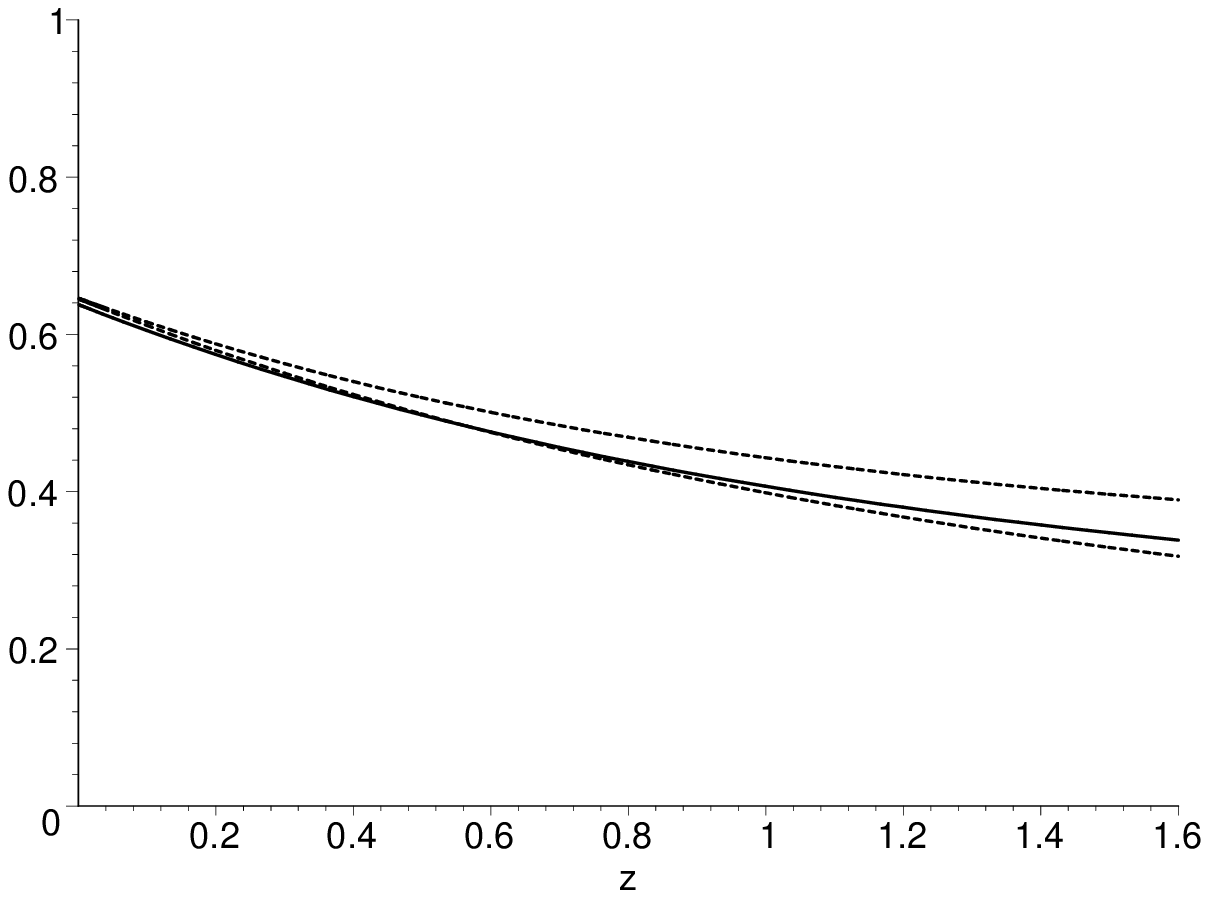}}}\smallskip
\centerline{{\bf(b)}\hskip-15pt
\scalebox{0.65}{\includegraphics{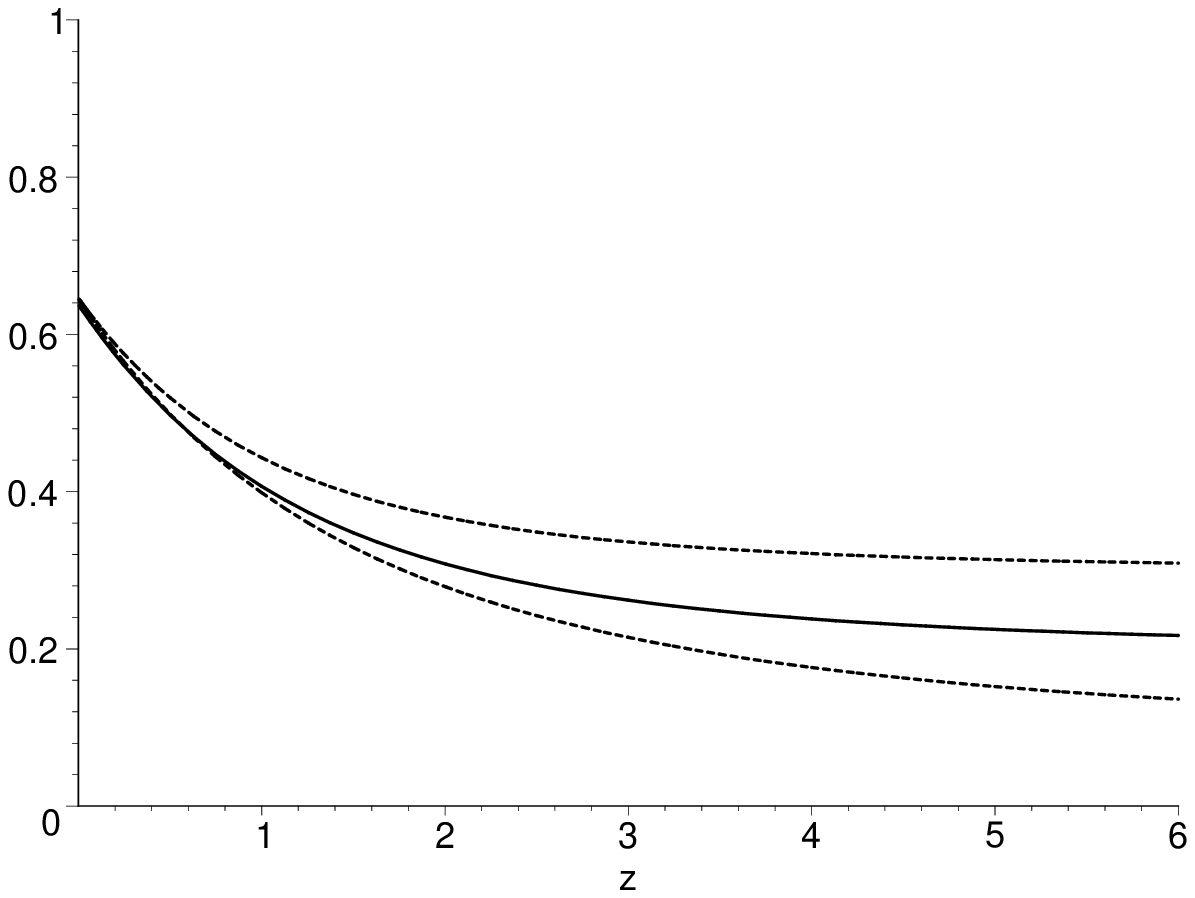}}}}
\def\figDv{\centerline{{\bf(a)}\hskip-15pt
\scalebox{0.75}{\includegraphics{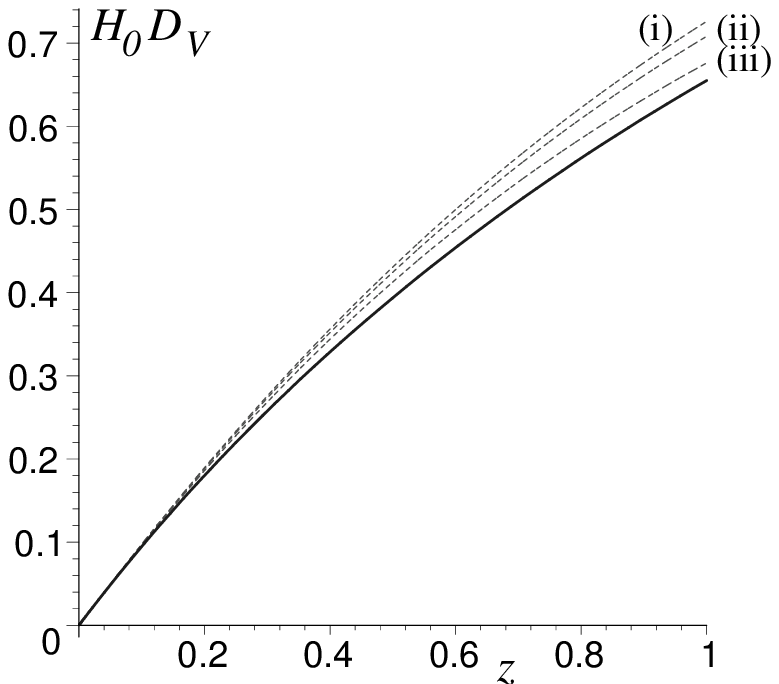}}}\smallskip
\centerline{{\bf(b)}\hskip-15pt
\scalebox{0.75}{\includegraphics{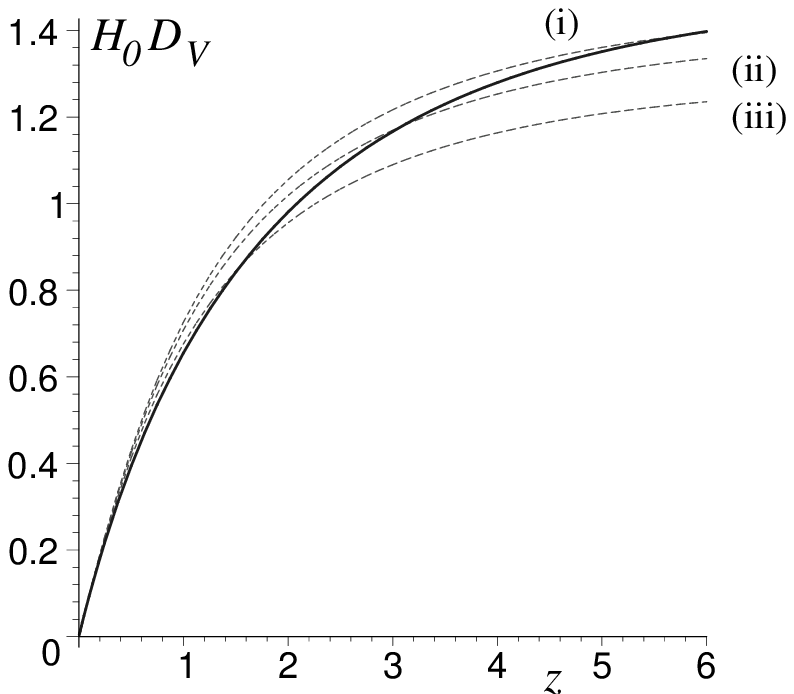}}}}
\begin{document}
%--------------------------------------------------------------------
\def\PRL#1{Phys.\ Rev.\ Lett.\ {\bf#1}} \def\PR#1{Phys.\ Rev.\ {\bf#1}}
\def\ApJ#1{Astrophys.\ J.\ {\bf#1}} \def\PL#1{Phys.\ Lett.\ {\bf#1}}
\def\AsJ#1{Astron.\ J.\ {\bf#1}} %{ApJ {\bf#1}}
\def\MNRAS#1{Mon.\ Not.\ R.\ Astr.\ Soc.\ {\bf#1}} %{MNRAS {\bf#1}}
\def\CQG#1{Class.\ Quantum Grav.\ {\bf#1}}
\def\GRG#1{Gen.\ Relativ.\ Grav.\ {\bf#1}} \def\Ei{E_\infty}
\def\AaA#1{Astron.\ Astrophys.\ {\bf#1}}
\def\beq{\begin{equation}} \def\eeq{\end{equation}}
\def\bea{\begin{eqnarray}} \def\eea{\end{eqnarray}}
\def\Z#1{_{\lower2pt\hbox{$\scriptstyle#1$}}} \def\w#1{\,\hbox{#1}}
\def\X#1{_{\lower2pt\hbox{$\scriptscriptstyle#1$}}} %\font\fiverm=cmr5
\font\sevenrm=cmr7 \def\ns#1{_{\hbox{\sevenrm #1}}} \def\dOM{\dd\Omega^2}
\def\Ns#1{\Z{\hbox{\sevenrm #1}}} \def\ave#1{\langle{#1}\rangle}
\def\lsim{\mathop{\hbox{${\lower3.8pt\hbox{$<$}}\atop{\raise0.2pt\hbox{$\sim$}}
$}}} \def\gsim{\mathop{\hbox{${\lower3.8pt\hbox{$>$}}\atop{\raise0.2pt\hbox{$
\sim$}}$}}} \def\kms{\w{km}\;\w{sec}^{-1}}\def\kmsMpc{\kms\w{Mpc}^{-1}}
\def\bn{\bar n} \def\dd{{\rm d}} \def\ds{\dd s} \def\etal{{\em et al}}
\def\al{\alpha}\def\be{\beta}\def\ga{\gamma}\def\de{\delta}\def\ep{\epsilon}
\def\et{\eta}\def\th{\theta}\def\ph{\phi}\def\rh{\rho}\def\si{\sigma}
\def\ta{\tau} \def\tn{\ts\Z0} \def\ac{a} \def\BB{{\cal B}} \def\CC{{\cal C}}
\def\frn#1#2{{\textstyle{#1\over#2}}} \def\Deriv#1#2#3{{#1#3\over#1#2}}
\def\Der#1#2{{#1\hphantom{#2}\over#1#2}} \def\pt{\partial} \def\ab{{\bar a}}
\def\goesas{\mathop{\sim}\limits} \def\tv{\ta\ns{v}} \def\tw{\ta\ns{w}}
\def\gb{\bar\ga}\def\omi{\OM_i} \def\I{{\hbox{$\scriptscriptstyle I$}}}
\def\av{{a\ns{v}\hskip-2pt}} \def\aw{{a\ns{w}\hskip-2.4pt}}\def\Vav{{\cal V}}
\def\DD{{\cal D}}\def\gd{{{}^3\!g}}\def\half{\frn12}\def\Rav{\ave{\cal R}}
\def\QQ{{\cal Q}}\def\dsp{\displaystyle} \def\rw{r\ns w}
\def\mean#1{{\vphantom{\tilde#1}\bar#1}} \def\bx{{\mathbf x}}
\def\bH{\mean H}\def\Hb{\bH\Z{\!0}} \def\gb{\mean\ga}\def\gc{\gb\Z0}
\def\rhb{\mean\rh}\def\OM{\mean\Omega}\def\etb{\mean\eta}
\def\fw{{f\ns w}}\def\fv{{f\ns v}} \def\goesas{\mathop{\sim}\limits}
\def\fvn{f\ns{v0}} \def\fvf{\left(1-\fv\right)} \def\Hh{H}
\def\OMM{\OM\Z M}\def\OMk{\OM\Z k}\def\OMQ{\OM\Z{\QQ}}
\def\epi{\epsilon_i}\def\gbi{\gb_i}\def\Omi{\OM\Z{Mi}}\def\OMkn{\OM_{k0}}
\def\Ci{C_\epsilon} \def\te{t_\epsilon}\def\OMMn{\OM\Z{M0}}
\def\OmMn{\Omega\Z{M0}} \def\OmBn{\Omega\Z{B0}} \def\OmCn{\Omega\Z{C0}} 
\def\ts{t} \def\tb{\ts'} \def\tc{\tau}
\def\la{\lambda}\def\Omb{\bar\Omega\X M} \def\dL{d\Z L} \def\dA{d\Z A} 
\def\name{timescape} \def\rhw{\rho\Z D} \def\Hm{H\Z0}
\def\Fi{\hbox{\footnotesize\it fi}}\def\etw{\eta\ns w} \def\etv{\eta\ns v}
\def\fvi{{f\ns{vi}}} \def\fwi{{f\ns{wi}}} \def\gw{\gb\ns w}\def\gv{\gb\ns v}
\def\Hv{H\ns v} \def\Hw{H\ns w} \def\kv{k\ns v} \def\hri{h_{ri}}
\def\LCDM{$\Lambda$CDM} \def\OmLn{\Omega\Z{\Lambda0}}\def\Omkn{\Omega\Z{k0}} 
\def\Dfb{D\Ns{TS}} \def\Dlcdm{D\Ns{$\scriptstyle\Lambda$CDM}}
\def\FF{{\cal F}} \def\HdA{\Hb\dA} \def\Aa{{\cal A}}
\def\tbA{\left(2t+3b\right)} \def\tbB{\left(2t^2+3bt+2b^2\right)}
\def\fdv{r\Z{0.35:0.2}}
%----------------------------------------------------------------
\title{ Average observational quantities in the timescape cosmology}
\author{David L. Wiltshire}
\email{David.Wiltshire@canterbury.ac.nz}
\affiliation{Department of Physics \& Astronomy, University of Canterbury,
Private Bag 4800, Christchurch 8140, New Zealand\footnote{Permanent address};}
\affiliation{International Center for Relativistic Astrophysics Network
(ICRANet), P.le della Repubblica 10, Pescara 65121, Italy}
%-----------------------------------------------------------------
\begin{abstract}
We examine the properties of a recently proposed observationally
viable alternative to homogeneous cosmology with smooth dark energy,
the timescape cosmology.
In the timescape model cosmic acceleration is realized as an apparent effect
related to the calibration of clocks and rods of observers
in bound systems relative to volume--average observers in an inhomogeneous
geometry in ordinary general relativity. The model is based on an exact
solution to a Buchert average of the Einstein equations with backreaction.
The present paper examines a number of observational tests which will
enable the timescape model to be distinguished from homogeneous
cosmologies with a cosmological constant or other smooth dark energy, in
current and future generations of dark energy experiments. Predictions are
presented for: comoving distance measures; $H(z)$; the equivalent of the
dark energy equation of state, $w(z)$; the $Om(z)$ measure of Sahni,
Shafieloo and Starobinsky; the Alcock--Paczy\'nski test; the baryon acoustic
oscillation measure, $D\Z V$; the inhomogeneity test of Clarkson, Bassett and
Lu; and the time drift of cosmological redshifts. Where possible, the
predictions are compared to recent independent studies of similar measures in
homogeneous cosmologies with dark energy. Three separate tests with
indications of results in possible tension with the \LCDM\ model are found to
be consistent with the expectations of the timescape cosmology.
\end{abstract}
%-----------------------------------------------------------------
\maketitle
%-----------------------------------------------------------------
\section{Introduction}
%-----------------------------------------------------------------
The paradigm for our current standard model of the universe assumes
that the universe is well--described by a geometry which is exactly
homogeneous and isotropic, with additional Newtonian perturbations.
The underlying geometry is assumed to be that of a
Friedmann--Lema\^{\i}tre--Robertson--Walker (FLRW) geometry, and in
matching the cosmological observables that derive from such a geometry,
we have been led to the conclusion over the past decade that the
present--day universe is dominated by a cosmological constant or other
fluid--like ``dark energy'' with an equation of state, $P=w\rho$, which
violates the strong energy condition.

Although the matter distribution was certainly very homogeneous at the epoch
of last--scattering when the cosmic microwave background (CMB) radiation was
laid down, however, in the intervening aeons the matter distribution has
become very inhomogeneous through the growth of structure. Large scale
surveys reveal the present epoch universe to possess a cosmic web of
structure, dominated in volume by voids, with galaxy clusters strung in
sheets and filaments that surround the voids, and thread them. Statistical
homogeneity of this structure appears only to be reached by averaging on
scales of order $100h^{-1}$ Mpc or more, where $h$ is the dimensionless
parameter related to the Hubble constant by $H\Z0=100h\kmsMpc$. The problem
of fitting a smooth geometry to a universe with such a lumpy matter
distribution \cite{fit1,fit2} is a nontrivial one, but central to
relating observations to the numerical values of the averaged parameters
which describe the Universe and its evolution as a whole.

Given the observed inhomogeneity of the present epoch universe, a number
of cosmologists have questioned whether the FLRW geometries are adequate
as a description of the universe at late times \cite{buch1}--\cite{EB}.
In particular, the
deduction that the universe is accelerating might in fact be a result of
trying to fit the wrong cosmological model. One central question in the
fitting problem is the issue of deriving the average evolution of the
inhomogeneous geometry. If one considers irrotational dust cosmologies,
and averages just inhomogeneous scalar quantities, in Buchert's scheme
\cite{buch1} one finds an average of the Einstein equation in which
there is a Friedmann--like evolution modified by backreaction
\cite{footK,footZ,footC},
\bea
&&\dsp{3\dot\ab^2\over\ab^2}=8\pi G\ave\rh-\half \Rav-\half\QQ,
\label{buche1}\\
&&{3\ddot\ab\over\ab}=-4\pi G\ave\rh+\QQ,\label{buche2}\\
&&\pt_t\ave\rh+3{\dot\ab\over\ab}\ave\rh=0,\label{buche3}\\
&&\pt_t\left(\ab^6\QQ\right)+\ab^4 \pt_t\left(\ab^2\Rav\right)=0,
\label{buche4}\eea
where an overdot denotes a time derivative for observers comoving with the
dust of density $\rh$, $\ab(t)\equiv\left[\Vav(t)/\Vav(t\Z0)\right]^{1/3}$
with $\Vav(t)\equiv\int_\DD\dd^3x\sqrt{\det\gd}$, angle brackets denote the
spatial volume average of a quantity, so that
$\Rav\equiv\left(\int_\DD\dd^3x\sqrt{\det\gd}\,{\cal R}(t,\bx)\right)/\Vav(t)$
is the average spatial curvature, and
\beq\QQ=\frn23\left(\langle\th^2\rangle-\langle\th\rangle^2\right)-
2\langle\si^2\rangle,\label{backr}\eeq
is the kinematic backreaction, $\si^2=\frn12\si_{\al\be}\si^{\al\be}$ being
the scalar shear. We use units in which $c=1$. Eq.\ (\ref{buche4}) is an
integrability condition needed to ensure that eq.~(\ref{buche1}) is the
integral of eq.~(\ref{buche2}).

One must be careful in interpreting equations (\ref{buche1})--(\ref{backr})
since the spatial averages refer to average quantities which depend on
the domain of integration on a spatial hypersurface. Observers measure
invariants of the local metric, not a spatially averaged metric, and
cosmological information comes to us on null geodesics. Given these
problems, the Buchert approach has been criticised \cite{IW}, and the
whole area of backreaction is the subject of some debate and controversy.
In recent work \cite {clocks,sol,essay} I have developed an interpretation
of solutions to the Buchert equations which circumvents the
criticisms of Sec.\ 3 of ref.\ \cite{IW}. It differs
from other approaches to the Buchert equations that have been used
in the literature \cite{morph}--\cite{RF}. As well as circumventing
objections that have been raised against Buchert averaging, the new
interpretation has a conceptual basis which can be understood as an
extension of the equivalence principle \cite{equiv}, and it leads to
a quantitative model universe with predictions \cite{clocks,LNW} which thus
far are in good agreement with observation. In particular, by Bayesian
comparison the Riess07 gold supernovae Ia (SneIa) data set \cite{Riess07}
agrees with the model predictions at a level which is
statistically indistinguishable from the standard spatially flat
\LCDM\ \cite{LNW,foot1}. The same best-fit parameters also fit
the angular scale of the sound horizon seen in CMB data, and the effective
comoving baryon acoustic oscillation (BAO) scale seen in angular diameter
tests of galaxy clustering statistics \cite{clocks,LNW}.

Given these promising indications, it is important that the
cosmology of refs.\ \cite{clocks,sol} is developed well beyond the stage 
of what might be regarded as a ``toy model'', so that it can be confronted
by all the same observational tests that are applied to the
\LCDM\ model. For example, many current precision tests involve
the detailed analysis of the CMB \cite{wmap}, and of galaxy clustering
statistics \cite{bao}--\cite{Seo}. To construct tests of
similar precision will require the development of new numerical
codes for the analysis of large datasets adapted to the present cosmology,
analogous to those based on the decades of detailed work that have been
applied to the standard cosmology.

Such goals represent an arduous project, and here I will simply take a few
steps in the direction of confronting the observations.
The aim of the present paper is not to present a detailed analysis of
current data sets, but to outline a number of observable quantities which
might be tested in future. Since the predictions obtained for a number of
these quantities can be readily compared to existing independent analyses
of homogeneous cosmologies with dark energy, I will make relevant comparisons
where possible. I will confine the
discussion here to average quantities which are relevant at all redshifts on
scales greater than the scale of statistical homogeneity. Other relatively
local tests which deal with quantities within the 100$h^{-1}$ Mpc scale of
statistical homogeneity \cite{clocks,dark07} will be left to future work.

The plan of the papers is as follows. In Sec.\ \ref{overview} I will
summarize the key features of the model introduced in refs.\
\cite{clocks,sol}, while also providing some further discussion. Additional
technical details which were not provided in ref.\ \cite{sol} on account of
space restrictions are given in the Appendices. In Sec.\ \ref{ldist} I discuss
the luminosity distance and angular diameter distance relations, and their
interpretation in terms of the equivalent of a ``dark energy equation
of state'', which enables a direct comparison to recent studies to be made.
In Sec.\ \ref{Hom} related diagnostics, $H(z)$ and the $Om(z)$ measure are
evaluated and discussed in relation to recent studies. The Alcock--Paczy\'nski
and BAO tests are treated similarly in Sec.\ \ref{APBAO}. The expected
nontrivial signature of a test of the Friedmann equation of Clarkson,
Bassett and Lu is determined in Sec.\ \ref{inhom}. A prediction
for the Sandage--Loeb test of the time drift of cosmological redshifts
is presented in Sec.\ \ref{drift}. Sec.\ \ref{dis} contains a concluding
discussion.

\section{Overview of the \name\ model\label{overview}}
\subsection{Voids and walls}

I will begin by briefly reviewing the two--scale ``fractal bubble model''
\cite{clocks} which I am hereby renaming the ``timescape model'' \cite{reason},
concentrating on the operational interpretation of observations. The model
is constructed by identifying the observed scales most relevant to the
observed present epoch inhomogeneous structure as being negatively curved
{\em voids} and spatially flat {\em walls}, which surround bound structures.

Since galaxies and galaxy clusters formed from perturbations which were
greater than critical density, then given an observable universe which
on average has negative Ricci scalar curvature, we have a natural
separation between walls and voids. As there is assumed to be a gradient
in spatial curvature, it is assumed we can always enclose the bound
structures which formed from over-critical perturbations within regions
which are spatially flat on average, and marginally expanding at the
boundary. These boundaries are called
{\em finite infinity} regions \cite{clocks}, with local average metric
\beq\ds^2\Z{\Fi}=-\dd\tw^2+\aw^2(\tw)\left[\dd\etw^2+
\etw^2\dOM\right]\,.
\label{wgeom}\eeq
The walls constitute the union of such finite infinity regions.
Observationally they would correspond to all extended structures that
contain galaxy clusters, namely sheets, filaments and knots \cite{sfk}.

Voids of a characteristic diameter $30h^{-1}$ Mpc are observed to fill
40--50\% of the universe at the present epoch \cite{HV}. In addition there are
numerous minivoids, which have been well--studied in the local volume
\cite{minivoids}. Together voids of all sizes appear to dominate the volume
of the present epoch universe, the exact volume fraction depending on the
empirical definition of a void in terms of some particular negative density
contrast. In the two--scale approximation of refs.\ \cite{clocks,sol}
both the dominant voids and minivoids are assumed to be characterized by the
same negatively spatial curvature scale, with a metric at the void centres
being given
by
\beq\ds^2\Z{\DD\ns{v}}=-\dd\tv^2+\av^2(\tv)\left[\dd\etv^2+\sinh^2(\etv)
\dOM\right]\label{vgeom}\eeq

We construct an average over the disjoint union of wall and void regions
over the entire present horizon volume
$\Vav=\Vav\ns i\ab^3$, where
\beq\ab^3=\fvi\av^3+\fwi\aw^3,\label{bav}\eeq
$\fvi$ and $\fwi=1-\fvi$
being the respective initial void and wall volume fractions at last
scattering. The finite infinity scale only becomes operationally defined once
regions start collapsing and structure forms. Thus at last scattering a
different interpretation of the wall and void components in (\ref{bav}) is
required. At this epoch the wall fraction $\fw$ is understood as that fraction
of the present horizon volume which comprises perturbations whose combined
mean density is the same as the mean density of the statistical ensemble
of perturbations, including those beyond the horizon. The void fraction,
$\fv$, is understood to be that (small) fraction of the present horizon
volume in underdense perturbations which was not compensated by overdense
perturbations at last scattering. It is convenient to rewrite (\ref{bav})
as $\fv(t)+\fw(t)=1$,
where $\fw(t)=\fwi\aw^3/\ab^3$ is the {\em wall volume fraction} and
$\fv(t)=\fvi\av^3/\ab^3$ is the {\em void volume fraction}.

\subsection{The scale of ``statistical homogeneity''}
In order to physically identify observables we must specify what is
to be identified as a ``particle'' of dust.
The question of what constitutes a particle of dust is not directly
addressed in Buchert's scheme, although perhaps implicitly many researchers
think of galaxies as being the particles of dust, as historically this
is the way the matter is treated in the FLRW model. However, galaxies
evolve considerably over time and are not homogeneously distributed at the
present epoch. Thus if we wish
to follow cosmic evolution from the epoch of last scattering to the
present, with no assumptions about homogeneity, then we must coarse grain
the dust on scales over which mass flows can be neglected, so that each
dust particle remains of a roughly fixed mass, even if the mass differs
somewhat from particle to particle.

Here we take a ``dust particle'' to be of at least the scale of
statistical homogeneity, 100$h^{-1}$ Mpc or somewhat larger \cite{footD}.
The {\em scale of statistical homogeneity} is taken to refer to a
scale volume within which the structure of voids and walls is roughly
similar, if such a box is chosen at random on a spatial slice in the
observable universe at late epochs. It is important to realise such volumes
will {\em not} have the same density. Rather they will have a density
which is distributed about a mean with a standard deviation of order several
percent, by an argument that follows from eq.~(\ref{est}) below.

We must stress that the universe is not considered to be a FLRW model,
and our terminology of a ``scale of statistical homogeneity'' is not
the same as in a FLRW model. A scale of homogeneity, in its sense in the
FLRW model, is not assumed to exist. The principal difference is that in a
FLRW model the density of the observable universe, sampled over the present
horizon volume, is assumed to be the mean density of the ensemble from
which our observable horizon volume was drawn. In the FLRW case
the standard deviation of the density of spatial volumes would decrease
to zero as one sampled ever larger volumes greater than the homogeneity
scale, as is the case for any stationary stochastic process \cite{statcos}.

Such a state of affairs cannot be expected to prevail, however,
given cosmic variance and an initial spectrum of density contrasts
of all possible length scales which are nested within each other,
which is the expectation from primordial inflation. Given cosmic
variance, then as one samples larger and larger volumes that become
comparable with the horizon volume, one is dealing with fewer and
fewer individual fluctuations rather than a statistical ensemble.
The assumption in the \name\ scenario is that the present horizon
volume is underdense relative to the ensemble mean density, which
at last--scattering is extremely close to critical.

The fact that it does nonetheless make sense to think of a ``scale of
statistical homogeneity'' as above, however, is a simple consequence of the
fact that although the density perturbations have all possible length
scales, the magnitude of these contrasts was strongly bounded at the
time of last scattering. In other words, given a universe which was close
to homogeneous at last scattering, it can only evolve so far from
homogeneity within the finite age of the universe.

The relevant scale for a cutoff to the ``scale of statistical homogeneity''
is scale of the largest acoustic wave in the plasma at last scattering -- of
order 110$h^{-1}$ Mpc. The simple reason for such a cutoff, is that below
this scale initial density contrasts may be amplified by acoustic waves in
the plasma, so that rather than having initial density contrasts of say
$\de\rh/\rh\goesas10^{-4}$ in nonbaryonic dark matter, the initial density
contrasts will be somewhat larger. The second acoustic peak in the CMB
anisotropy spectrum -- i.e., the first refraction peak -- for example, will
amplify the density contrast of underdense regions, and may therefore be the
feature of the primordial spectrum responsible for the fact that the dominant
void fraction is associated to a specific scale $30h^{-1}$ Mpc \cite{HV}.

Given that some initial density perturbations are amplified below the
acoustic scale, and that the CMB anisotropy spectrum is fairly flat
at long wavelengths, the acoustic scale provides a cutoff analogous the
cutoff between the nonlinear and linear regimes of structure formation,
although here there is no single global FLRW model about which such
regimes linear are defined \cite{footL}. Below the scale of statistical
homogeneity we will typically find density contrasts $|\de\rh/\rh|
\goesas1$ which characterize the nonlinear regime, as is the observed
case for $30h^{-1}$ Mpc diameter voids \cite{HV}. Above the acoustic scale,
we can be sure that the perturbations at last scattering have very similar
amplitudes as a function of scale. Although the perturbations
in the photon--baryon plasma have contrasts $\de\rh/\rh\goesas10^{-5}$
at this epoch, the density contrast in nonbaryonic dark matter is expected
to be somewhat larger; e.g., of order $\de\rh/\rh\goesas10^{-4}$--$10^{-3}$,
depending on one's dark matter model.

The standard deviation of the density of cells on scales larger than the
scale of statistical homogeneity can be estimated crudely by assuming
that such cells each evolve as an independent Friedmann universe from a smooth
perturbation at the epoch of last scattering. This approximation is
justified since the relevant scale is the one over which there are no
appreciable average mass flows from one dust cell to another. We assume that
the backreaction contributions do not dominate the volume--average evolution,
and make our rough estimate from the Friedmann equation with pressureless
dust only, for which
$$a\Z0^2\Hm^2(\OmMn-1)=a^2(t)H^2(t)[\Omega\Z M(t)-1]$$
This leads to a present epoch density contrast
\beq
\de\rh\Z0\simeq\left(H\over\Hm\right)^2{\de\rh_t\over(1+z)^2}\,,
\label{est}\eeq
where the density contrast is relative to the critical density, so that
$\de\rh_t=\Omega\Z M(t)-1$ etc, where $\Omega\Z M$ is a density parameter
for the isolated region only. (Physically, the critical density is that
within a spatially flat wall region.)
Thus if we take $\de\rh_t\simeq10^{-4}$ at last scattering, when $z\simeq1090$
and when $H\simeq 2/(3t)$ with $t\simeq380,000\w{yr}$, we are led to
$\de\rh\Z0\simeq0.025/h^2\simeq0.06$ if $h\simeq0.65$.

This crude estimate can be compared to the actual density variance determined
from large scale structure surveys \cite{Hogg,Sylos}. Sylos Labini \etal\
\cite{Sylos} have recently
determined the variance in the number density of luminous red galaxies (LRGs)
in the SDSS-DR7 by dividing the full sample of 53,066 galaxies in the redshift
range $10^{-4}<z <0.3$ into $N$ equal nonoverlapping volumes. Over the range
$4\le N\le15$, the standard deviation is found to be of order 8\%, consistent
with an earlier measurement of 7\% by Hogg \etal\ \cite{Hogg} in a smaller LRG
sample. These values are very close to our order of magnitude estimate of 6\%.
Provided LRGs are correlated to
the actual density, then the variance in the percentage density contrast will
be commensurate. In fact, such variances
can be used to constrain the dark matter density contrast at last scattering.
A measurement of 8\% would indicate, reversing the argument above, that
a contrast of $\de\rh/\rh\goesas10^{-3}$ in nonbaryonic dark matter at
last scattering is an order of magnitude too large.

Given a nearly scale--invariant spectrum of density perturbations, with
perturbations nested in perturbations, our expectation is that the variance
in density should not decrease appreciably if sample volumes are increased
at nearby redshifts. In principle, it should be possible to calculate it
as a function of scale, given the constraints from the CMB anisotropy
spectrum at long wavelengths. For spatial slices at higher redshifts, looking
further back in time, the variance would decrease in accord with (\ref{est})
-- provided that a sample of objects such as LRGs can be found which
does not exhibit strong evolutionary effects over the range of redshifts in
question.

Of course, the estimate based on (\ref{est}) could be further
refined to take backreaction into account; but further accuracy can only
be gained when one has a tighter estimate of the dark matter density
contrast than simply an order of magnitude. Furthermore, the statistical
physics of cosmic structures in the \name\ scenario may well differ
from that of the FLRW model \cite{statcos} in significant ways; one has to
revisit the whole problem from first principles.

In summary, the observed universe is not assumed to be homogeneous or
to approach any single global FLRW model at any scale. The ``statistical
homogeneity scale'' -- which will coincide roughly with the BAO scale
-- represents a scale above which the variance in density contrasts is
bounded at the 10\% level, and below which density constrasts become
as large as they can possibly be.

\subsection{The bare Hubble flow and bare cosmological parameters}

Given our identification of dust particles coarse-grained at the scale
of statistical homogeneity, and possible very large differences in
spatial curvature and gravitational energy within such a cell, we
do not assume that Buchert average time parameter, $t$, is the relevant
parameter measured by every isotropic observer --  those who see an
isotropic CMB -- within any dust cell. Rather it is the time parameter
measured by an isotropic observer whose local spatial curvature happens
to coincide with the Buchert volume--average spatial curvature $\Rav$.
We employ an ansatz of an underlying quasilocal uniform Hubble flow
within a dust cell, below the scale of statistical homogeneity, in
terms of local proper lengths with respect to local proper times, which
both vary with gradients in spatial curvature and gravitational energy.
This ansatz provides an implicit resolution of the Sandage--de Vaucouleurs
paradox \cite{clocks}, and can be understood in terms of a generalization
of the equivalence principle \cite{equiv}.

The metrics (\ref{wgeom}) and (\ref{vgeom}) are assumed to
represent the local geometry for isotropic observers at finite infinity and
at void centres respectively. Within the scale of statistical homogeneity the
metrics (\ref{wgeom}) and (\ref{vgeom}) are assumed to be patched together
with a condition of uniform quasilocal bare Hubble flow \cite{clocks,equiv}
\beq
\bH= \Deriv\dd\tw\aw=\Deriv\dd\tv\av,
\eeq
which will preserve isotropy of the CMB. The mean CMB temperature and angular
anisotropy scale will vary with the gradients in gravitational energy and
spatial curvature, however.

For the purpose of the Buchert average we refer all quantities to one set of
volume--average clocks: those that keep the time parameter
$\ts$ of eqs.\ (\ref{buche1})--(\ref{backr}) so that
\beq
\bH\equiv{\dot\ab\over\ab}=\gw\Hw=\gv\Hv
\eeq
where
\beq\Hw\equiv{1\over\aw}\Deriv\dd t\aw,\qquad\w{and}\qquad
\Hv\equiv{1\over\av}\Deriv\dd t\av\,,\label{homo3}\eeq
and
\beq\gw\equiv\Deriv\dd\tw{t\ },\qquad\w{and}\qquad\gv=\Deriv\dd\tv{t\ },
\label{clocks1}
\eeq
are lapse functions of volume--average time, $t$, relative to wall and
void--centre observers respectively. The ratio of the relative Hubble
rates $h_r=\Hw/\Hv<1$ is related to the wall lapse function by
\beq\gw=1+{(1-h_r)f_v\over h_r},\label{clocks2}\eeq
and $\gv=h_r\gw$.

The Buchert equations for pressureless dust with volume--average density
$\rhb\Z M$ are solved \cite{sol} in the two--scale approximation
by assuming that there is no backreaction within walls and voids separately
\cite{foot2}, but only in the combined average. With this assumption, the
kinematic backreaction term becomes \cite{clocks}
\beq\QQ=6\fv(1-\fv)\left(\Hv-\Hw\right)^2=
{2\dot\fv^2\over3\fv(1-\fv)}\,.\label{Q1}
\eeq
The resulting independent Buchert equations consist of two coupled nonlinear
ordinary differential equations \cite{clocks} for $\ab(t)$ and $\fv(t)$,
which may be written as
\bea
&&\OMM+\OMk+\OMQ=1,\label{Beq1}\\
&&\ab^{-6}\pt_t\left(\OMQ\bH^2\ab^6\right)+\ab^{-2}\pt_t\left(\OMk\bH^2\ab^2
\right)=0\,,\label{Beq2}
\eea
where
\bea\OMM&=&{8\pi G\rhb\Z{M0}\ab\Z0^3\over 3\bH^2\ab^3}\,,\label{omm}\\
\OMk&=&{-\kv\fvi^{2/3}\fv^{1/3}\over \ab^2\bH^2}\,,\label{omk}\\
\OMQ&=&{-\dot\fv^2\over 9\fv(1-\fv)\bH^2}\,,\label{omq}
\eea
are the volume--average or ``bare'' matter density, curvature density and
kinematic backreaction density parameters respectively, $\ab\Z0$
and $\rhb\Z{M0}$ being the present epoch values of $\ab$ and $\rhb\Z M$.
The average curvature is due to the voids only, which are assumed
to have $\kv<0$. The volume--average deceleration parameter is given by
\beq
\bar q\equiv{-\ddot\ab\over\bH^2\ab}=\half\OMM+2\OMQ\label{vdec}
\eeq

Equations (\ref{Beq1}), (\ref{Beq2})
are readily integrated to yield an exact solution \cite{sol}, which is
listed in Appendix \ref{apps}, together with its simple tracking limit in
Appendix \ref{appt}. For initial conditions at last scattering
consistent with observations, solutions are found to reach within 1\%
of the tracking limit by a redshift $z\goesas37$ \cite{sol}. Thus the
tracker solution will be used for the purposes of the specific cosmological
tests which are investigated in this paper \cite{footT}.

It should be noted that for the solution found in ref.\ \cite{sol}, the
backreaction term is at most of order $4.2\%$ \cite{footQ}. Its redshift
dependence for the best--fit parameters is exhibited in Fig.~\ref{fig_omq}.
Although $\OMQ$ is negative, it is never large enough relative to $\OMM$ to
dominate the r.h.s.\ of (\ref{vdec}) and give volume average cosmic
acceleration.
The backreaction itself is not the sole reason for apparent cosmic
acceleration; that is also a question of how volume--average
evolution is interpreted in terms of a local metric.
%------------------------------------------------------------
\begin{figure}[htb]
\vbox{\figomq
\caption{\label{fig_omq}%
{\sl The bare backreaction density parameter $\bar\Omega\Z{\cal Q}$ as a
function of redshift for the \name\ model with $\fvn=0.762$,
$\Hm=61.7\kmsMpc$.}}}
\end{figure}
%------------------------------------------------------------

\subsection{Dressed cosmological parameters}
One must take care in physically interpreting the solution of the Buchert
equation, since it does not represent a single exact solution of
Einstein's equations, but rather a spatial average. Observers measure
invariants of the local metric, and information carried by radial null
geodesics from distant parts of the universe. In ref.\ \cite{clocks}
a means of interpreting the Buchert equation was developed as follows.

First, since cosmological information is obtained by a radial spherically
symmetric average, we construct a spherically symmetric geometry relative
to an observer who measures volume--average time, and with a spatial
volume scaling as $\ab^3(t)$,
\beq
\dd\mean s^2=-\dd\ts^2+\ab^2(\ts)\,\dd\etb^2+\Aa(\etb,\ts)\,\dOM,\nonumber\\
\label{avgeom}
\eeq
where the area quantity, $\Aa(\etb,\ts)$, satisfies
$\int^{\etb\X{\cal H}}_0\dd\etb\, \Aa(\etb,\ts)=\ab^2(\ts)\Vav\ns{i}
(\etb\Z{\cal H})/(4\pi)$, $\etb\Z{\cal H}$ being the conformal distance to
the particle horizon relative to an observer at $\etb=0$, since we
have chosen the particle horizon as the scale of averaging. The metric
(\ref{avgeom}) is spherically symmetric by construction, but is not a
Lema\^{\i}tre--Tolman--Bondi (LTB) solution since it is not an exact solution
of Einstein's equations, but rather of the Buchert average of the Einstein
equations.

In terms of the wall time, $\tw$, of finite infinity observers
the metric (\ref{avgeom}) is
\beq\dd\mean s^2=-\gw^2(\tw)\dd\tw^2+\ab^2(\tw)\,\dd\etb^2+\Aa(\etb,\tw)\,\dOM
\,.\label{avgeom2}\eeq
However, this geometry, which has negative spatial curvature is not the
locally measured geometry at finite infinity, which is given instead
by (\ref{wgeom}). Since (\ref{wgeom}) is not a global geometry, we
match (\ref{wgeom}) to (\ref{avgeom2}) to obtain a dressed wall geometry,
which is effectively the closest thing there is to a FLRW geometry
adapted to the rods and clocks of wall observers. The matching is achieved
in two steps. First we conformally match radial null geodesics of
(\ref{wgeom}) and (\ref{avgeom2}), bearing in mind that null geodesics are
unaffected by an overall conformal scaling. This leads to a relation
\beq
\dd\etw={\fwi^{1/3}\dd\etb\over\gw\fvf^{1/3}}\label{etarel}
\eeq
along the geodesics. Second, we account for volume and area factors by taking
$\etw$ in (\ref{wgeom}) to be given by the integral of (\ref{etarel}).

The wall geometry (\ref{wgeom}), which may also be written
\beq \ds^2\Z{\Fi}=-\dd\tw^2+{\fvf^{2/3}\ab^2\over\fwi^{2/3}}
\left[\dd\etw^2+\etw^2\dOM\right]\,,
\eeq
on account of (\ref{bav}), is a local geometry only valid in spatially flat
wall regions. We now use (\ref{etarel}) and its integral to extend this
metric beyond the wall regions to obtain the dressed global metric
\bea
\ds^2&=&-\dd\tw^2+{\ab^2\over\gw^2}\,\dd\etb^2+
{\ab^2\fvf^{2/3}\over\fwi^{2/3}}\,\etw^2(\etb,\tw)\,\dOM\nonumber\\
&=&-\dd\tw^2+\ac^2(\tw)\left[\dd\etb^2+\rw^2(\etb,\tw)\,\dOM\right]
\label{dgeom}\eea
where $a\equiv\gw^{-1}\ab$, and 
$$\rw\equiv\gw\fvf^{1/3}\fwi^{-1/3}\etw(\etb,\tw).$$ Whereas (\ref{wgeom})
represents a local geometry only valid in spatially flat wall regions,
the dressed geometry (\ref{dgeom}) extends as an average effective geometry
\cite{foot5} to the cosmological scales
parameterized by the volume--average conformal time, which satisfies
$\dd\etb=\dd t/\ab=\dd\tw/ \ac$. Since
the geometry on cosmological scales does not have constant Gaussian curvature
the average metric (\ref{dgeom}), like (\ref{avgeom}), is spherically symmetric
but not homogeneous.

In trying to fit a FLRW model to the universe, the cosmological
parameters we obtain effectively have numerical values close to those of the
dressed geometry (\ref{dgeom}). In particular, we infer a dressed
matter density parameter
\beq \Omega\Z{M}=\gw^3\OMM\,,\eeq
a dressed Hubble parameter
\bea\Hh\equiv{1\over\ac}\Deriv\dd\tw\ac
={1\over\ab}\Deriv\dd\tw\ab-{1\over\gw}\Deriv\dd\tw\gw
=\gw\bH-\dot\gw\,,\label{42}
\eea 
and similarly a dressed deceleration parameter, where the overdot still
denotes a derivate with respect to volume--average time. As demonstrated
in refs.\ \cite{clocks,sol} in a void--dominated universe the dressed
deceleration parameter is negative at late epochs, even though the bare
deceleration parameter (\ref{vdec}) is positive. Thus cosmic acceleration
is realised as an apparent effect due to the variance of local geometry from
the average, leading to variance in the calibration of clocks and rods.

In the rest of the paper we will drop the subscript ``w'' from both $\tw$
and $\gw$, as we will not need to
make explicit reference to the time measured in
void centres. Thus $\tc$ and $\gb$ will be assumed to refer to wall time.
\medskip

\section{Comoving distance $D(z)$ and equivalent of the ``equation of state''%
\label{ldist}}

In testing fluid--like dark energy scenarios, or modified gravity theories
that can be cast as an effective fluid with an equation of state $P=w\rho$,
a common question is: how can the equation of state parameter, $w(z)$,
be constrained as a function of redshift? Unfortunately, if dark energy
is some purely unknown physics, then it is completely unclear how one
should expand it as a power series. A linear series in $z$ will not converge
for $z>1$, for example, so series in $z/(1+z)$ are sometimes considered.
Unless one has a precise physical model of dark energy to be tested,
then any constraints are completely dependent on how one chooses to
characterize such a power series.
When constraints on the value of $w$ from cosmological observations
are quoted in the literature, it is often on the basis
that $w$ is simply a constant, even though there is no known
physics for making such an assumption, apart from the cosmological constant
case of $w=-1$.

In this section I will derive the equivalent of the ``equation of state''
style observational tests, although the
terminology ``equation of state'' does not have a meaning in terms
of actual observables, given that the model in question is not characterized
by a fluid with $P\Z D=w\rhw$. Let us recall that in the case of the standard
FLRW models, the equation of continuity for such a dark energy component in a
background universe with scale factor $a(t)$, viz.,
\beq
\dot\rhw+3{\dot a\over a}(1+w)\rhw=0\,,
\eeq
may be integrated to give
\beq
\ln\left(\rhw\over\rh\Z{D0}\right)=\int{3[1+w(z)]\dd z\over 1+z}\,
\label{rhw}\eeq
using $a\Z0/a=1+z$, where it is assumed that the equation of state parameter
varies with redshift. To obtain an expression for the luminosity distance
one substitutes (\ref{rhw}) in the spatially flat Friedmann equation for
matter plus dark energy,
\beq
{\dot a^2\over a^2}={8\pi G\over3}\left[\rh\Z{M0}\left(a\Z0\over a\right)^3
+\rhw\right]
\label{edark}\eeq
and uses the resulting expression for $\dot a$, to determine the
conformal time integral
\begin{widetext}
\beq
r\Ns{FLRW}\equiv\int_\ts^{\ts\X0}{\dd t\over a}
=\int_a^{a\X0}{\dd a\over a\dot a}\\
=\int_0^z{\dd z'\over a\Z0\Hm\sqrt{\OmMn(1+z')^3+\Omega\Z{D0}
\exp\left[3\int_0^{z'}{(1+w(z''))\dd z''\over 1+z''}\right]}}\,,
\label{rFLRW}\eeq
\end{widetext}
where $\OmMn=8\pi G\rh\Z{M0}/(3\Hm^2)$ and $\Omega\Z{D0}=8\pi G\rh\Z{D0}/
(3\Hm^2)=1-\OmMn$. The standard luminosity distance is then given by
$\dL=a\Z0 r\Ns{FLRW}(1+z)$. The quantity
\beq
D=a\Z0 r\Ns{FLRW}={\dL\over1+z}
\label{DFLRW}\eeq
is the comoving distance quantity directly related to the luminosity distance.
The angular diameter distance is also related by
\beq\dA={D\over1+z}={\dL\over(1+z)^2}\,.\label{dist}\eeq
We observe from (\ref{rFLRW}) that $\Hm D$ does not depend on the value
of the Hubble constant, $\Hm$, but only directly on $\OmMn$.

Given observed quantities such as the apparent luminosity--redshift
relation or an angular size--redshift relation for standard candles
or standard rulers, we can take derivatives of (\ref{rFLRW}) to obtain
\beq
w(z)={\frn23(1+z)D'^{-1}D''+1\over\OmMn(1+z)^3\Hm^2 D'^2-1}
\label{eos}\eeq
where the prime denotes a derivative with respect to $z$.
This gives a formal equation of state to any comoving distance relation,
assuming an underlying spatially flat dark energy model. Such a relation
can be applied to observed distance measurements, regardless of whether
the underlying cosmology has dark energy or not. We should note, however,
that such a $w(z)$ has first and second derivatives of the observed
quantities, and so is much more difficult to determine observationally
than direct fits to a quantity such as $D(z)$.

For the \name\ universe, equivalent comoving, angular diameter and luminosity
distances can be defined in terms of the dressed geometry (\ref{wgeom}).
We have a dressed luminosity distance relation 
\beq\dL=a\Z0(1+z)\rw,\eeq where $a\Z0=\gc^{-1}\ab\Z0$, and the
{\em effective comoving distance} to a redshift $z$ is $D=a\Z0\rw$, where
\beq\rw=\gb\fvf^{1/3}
\int_\ts^{\ts\X0}{\dd\tb\over\gb(\tb)(1-\fv(\tb))^{1/3}\ab(\tb)}\,.
\label{eq:dL}\eeq
As discussed in Sec.\ \ref{overview}, since spatial sections are not of
constant Gaussian curvature this
effective comoving distance represents a fit to our spatially flat rods
once radial null geodesics are conformally matched, and geometric factors
are taken into account.

For the tracker solution (\ref{track1}), (\ref{track2}) the cosmological
redshift satisfies
\beq
z+1={\ab\Z0\gb\over\ab\gc}={(2+\fv)\fv^{1/3}\over3\fvn^{1/3}\Hb t}
={2^{4/3}t^{1/3}(t+b)\over\fvn^{1/3}\Hb t(2t+3b)^{4/3}}\,,
\label{redshift}\eeq
where
\beq b={2(1-\fvn)(2+\fvn)\over9\fvn\Hb}\,.\label{bb}\eeq
The integral in (\ref{eq:dL}) is readily evaluated to give
\bea
\dA&=&{D\over 1+z}=\ts^{2/3}\int_\ts^{\ts\X0}
{2\,\dd\tb\over(2+\fv(\tb))(\tb)^{2/3}}\nonumber\\
%=y^2\left[2y+{b\over6}\ln\left((y+b)^2\over y^2-by+b^2\right)
%+{b\over\sqrt{3}}\tan^{-1}\left(2y-b\over\sqrt{3}\,b\right)\right]^{y\X0}_y
&=&{t^{2/3}(\FF(t\Z0)-\FF(t))}
\label{lumd}
\eea
where
\bea
\FF(t)&=&2t^{1/3}+{b^{1/3}\over6}\ln\left((t^{1/3}+b^{1/3})^2\over
t^{2/3}-b^{1/3}t^{1/3}+b^{2/3}\right)\nonumber\\
&&\qquad+{b^{1/3}\over\sqrt{3}}\tan^{-1}\left(2t^{1/3}-b^{1/3}\over
\sqrt{3}\,b^{1/3}\right).\label{FF}
%\Hb\dA&=&{\Hb\dL\over(1+z)^2}=\left(\Hb\ts\right)^{2/3}\int_\ts^{\ts\X0}
%{2\Hb\dd\tb\over(2+\fv(\tb))(\Hb\tb)^{2/3}}\nonumber\\
%&=&y^2\left[2y+{b\over6}\ln\left((y+b)^2\over y^2-by+b^2\right)
%+{b\over\sqrt{3}}\tan^{-1}\left(2y-b\over\sqrt{3}\,b\right)\right]^{y\X0}_y
\eea

It is straightforward now to compare distance measurements in the
\name\ model with those in spatially flat \LCDM\ models. The
\name\ model which best fits the Riess07 gold data set had a void fraction
$\fvn=0.76^{+0.12}_{-0.09}$, and dressed Hubble constant
$\Hm=61.7^{+1.2}_{-1.1}\kmsMpc$, where $1\si$ uncertainties are quoted
\cite{LNW}. In Fig.~\ref{fig_coD} we plot
\beq
\Hm D=\Hm t^{2/3}[\FF(\tn)-\FF(t)](1+z)
\eeq
for the best--fit model with $\fvn=0.762$, as compared to three spatially
flat \LCDM\ models with different values of $\OmMn$, (or of $\OmLn=1-\OmMn$).
Fig.~\ref{fig_coD} shows that over redshifts between the present epoch and
last scattering, the \name\ model interpolates between
\LCDM\ models with different values of $\OmMn$. For redshifts $z\lsim1.5$
$\Dfb$ is very close to $\Dlcdm$ for the parameter values $(\OmMn,\OmLn)
=(0.34,0.66)$ (model (iii)) which best--fit the Riess07 SneIa data
only \cite{LNW}. For very large
redshifts that approach the surface of last scattering, $z\lsim1100$, on the
other hand, $\Dfb$ very closely matches $\Dlcdm$  for the parameter values
$(\OmMn,\OmLn) =(0.249,0.751)$ (model (i)) which best--fit WMAP5 only
\cite{wmap}. Over redshifts $2\lsim z\lsim10$, at which scales independent
tests are conceivable, $\Dfb$ makes a transition over corresponding
curves of $\Dlcdm$ with intermediate values of $(\OmMn,\OmLn)$. The
$\Dlcdm$ curve for joint best fit parameters to SneIa, BAO measurements and
WMAP5 \cite{wmap}, $(\OmMn,\OmLn) =(0.279,0.721)$ is best--matched over
the range $5\lsim z\lsim 6$, for example.

Given the difference of $\Dfb$ from any single $\Dlcdm$ curve
becomes pronounced only in the range $2\lsim z\lsim 6$, it may be difficult
to distinguish the models on the basis of the measurement of $\dA$ alone
from BAO surveys, which will be able to measure $\dA(z)$ up to 1\% to $z<3$. 
However, joint measurements of other parameters, such as $H(z)$, may
make for definitive tests, as will be discussed later. Gamma--ray bursters
(GRBs) do probe distances to redshifts $z\lsim8.3$, and could be very useful.
There has already been much work deriving Hubble diagrams using GRBs. (See,
e.g., \cite{GRB}.) It would appear that more work needs to be done
to nail down systematic uncertainties, but GRBs may provide a definitive
test in future. An analysis of the \name\ model Hubble diagram using 69 GRBs
has just been performed by Schaefer \cite{Schaefer}, who finds that it fits
the data better than the concordance \LCDM\ model, but not yet by a
huge margin. As more data is accumulated, it should become possible to
distinguish the models.

\subsection{Effective ``dark energy equation of state''}
The equivalent of an equation of state, $w(z)$, for the \name\ model
may be determined from (\ref{eos}) and (\ref{lumd}). The specific
analytic expressions for the first and second derivatives of $D$ are
\beq
\Deriv\dd z D={t\left[\left(2t-b\right)\dA+\tbA^2\right]
\over3\tbB}\,,\label{dDdz}
\eeq
\begin{widetext}
${}$
%------------------------------------------------------------
\begin{figure}[htb]
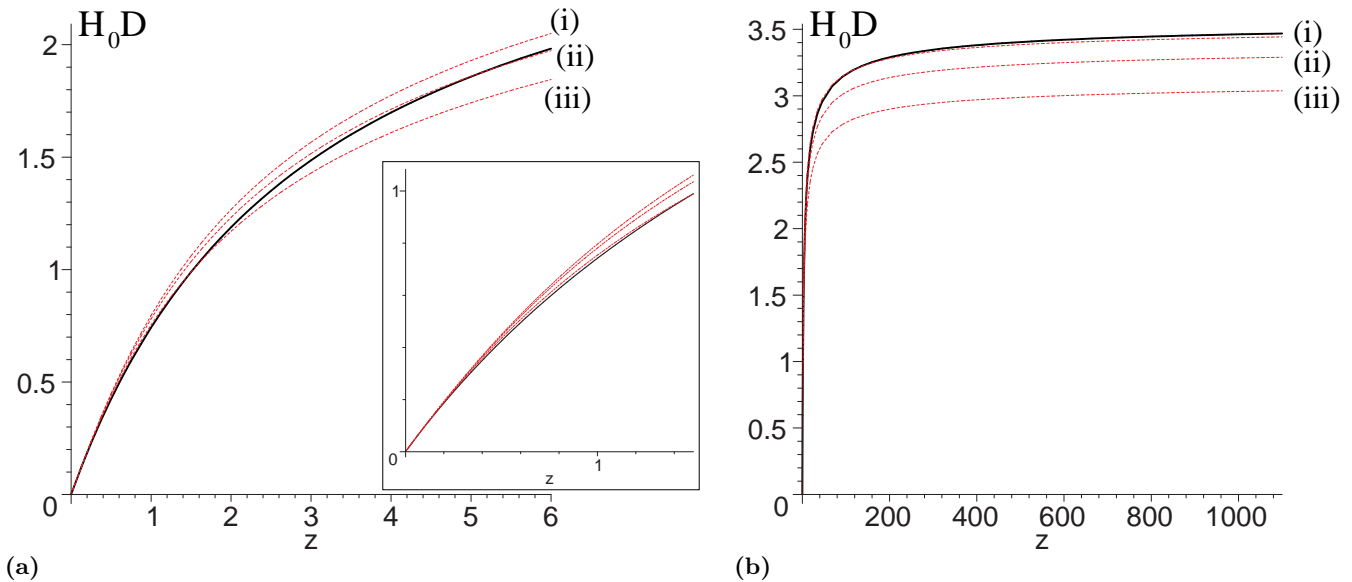

\vbox{\figcoD %\figcoDa \figcoDb
\caption{\label{fig_coD}%
{\sl The effective comoving distance $\Hm D(z)$ is plotted for
the best--fit \name\ model, with $\fvn=0.762$, (solid line); and
for various spatially flat \LCDM\ models (dashed red lines). The parameters for
the dashed lines are (i) $\OmMn=0.249$ (best--fit to WMAP5 only); (ii)
$\OmMn=0.279$ (joint best--fit to SneIa, BAO and WMAP5); (iii)  $\OmMn=0.34$
(best--fit to Riess07 SneIa only). Panel {(a)} shows the redshift range $z<6$,
with an inset for $z<1.5$, which is the range tested by current SneIa data.
Panel {(b)} shows the range $z<1100$ up to the surface of last scattering,
tested by WMAP5.}}}
\end{figure}
%------------------------------------------------------------
%\end{widetext}
\beq
{\dd^2D\over\dd z^2}=
{-t\tbA\left[2\left(t+b\right)\left(2t+5b\right)\left(2t^2+3bt-b^2\right)
\dA+\tbA\left(8t^4+26bt^3+53b^2t^2+56b^3t+18b^4\right)\right]
\over9(1+z)\tbB^3}
\label{dDdz2}\eeq
In these expressions $\dA$ is given by (\ref{lumd}) and $t$ is
given implicitly in terms of the redshift, $z$, via (\ref{redshift}).
We now substitute (\ref{dDdz}) and (\ref{dDdz2}) in (\ref{eos}) and
use the fact that by (\ref{HbH}), $\Hm=(4\fvn^2+\fvn+4)\Hb/[2(2+\fvn)]$,
to obtain
\beq
w={\left(40t^5-28bt^4-274b^2t^3-349b^3t^2-92b^4t+24b^5\right)
\dA+t\tbA^2\left(20t^3+56bt^2+47b^2t-4b^3\right)\over
\left\{\left(2t-b\right)\dA+\tbA^2\right\}\left\{A\Z0(z+1)^3t^2
\left[(2t-b)\dA+\tbA^2\right]^2-9\tbB^2\right\}}
\label{eosTS}\eeq
\end{widetext}
where
\beq
A\Z0={\OmMn(4\fvn^2+\fvn+4)^2\over4(2+\fvn)^2}
\eeq
and $\dA$ is given by (\ref{bb}) and (\ref{lumd}). In fact, $\OmMn=$\break
$\frn12(1-\fvn)(2+\fvn)$, so that
$$A\Z0={(1-\fvn)(4\fvn^2+\fvn+4)^2\over8(2+\fvn)}.$$
%------------------------------------------------------------
\begin{figure}[htb]
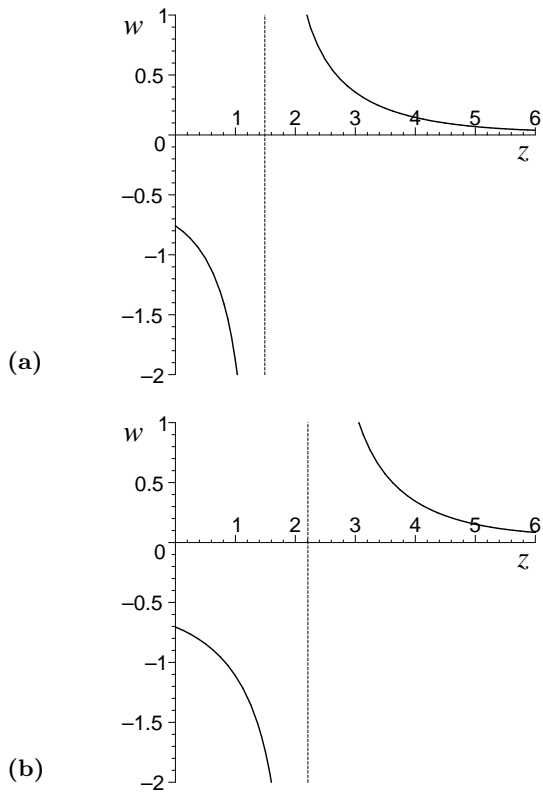

\vbox{\figwza\figwzb
\caption{\label{fig_wz}%
{\sl The artificial equivalent of an equation of state (\ref{eosTS}),
constructed using the effective comoving distance
(\ref{eos}), plotted for the \name\ tracker solution with best--fit value
$\fvn=0.762$, and two different values of $\OmMn$: {\bf(a)} the canonical
dressed value $\OmMn=\frn12(1-\fvn)(2+\fvn)=0.33$; {\bf(b)} $\OmMn=0.279$.}}}
\end{figure}
%------------------------------------------------------------

Since the $w(z)$ expression is an artificial mathematical
construction for the present model, we can also determine $w(z)$
if a value of $\OmMn$ different from the canonical value
$\frn12(1-\fvn)(2+\fvn)$ is assumed. In this way,
we arrive at the example $w(z)$ curves plotted in Fig.~\ref{fig_wz}.
The fact that the denominator of (\ref{eosTS}) goes through zero
means that $w(z)$ becomes formally infinite and changes sign at a
value of $z$ which depends on the value of $\OmMn$ assumed \cite{footN}. This
feature illustrates how pointless it is to talk about an equation
of state of dark energy, or to choose to ``reconstruct'' $w(z)$ if
the underlying unknown physics has nothing to do with a fluid in the vacuum
of space. What is actually measured is a quantity such as $D(z)$,
illustrated in Fig.~\ref{fig_coD}, and this is perfectly smooth.

Phenomenologically, for the canonical best--fit dressed value of $\OmMn=0.33$
\cite{LNW}, one finds that $w(0)\simeq-0.758$ and that $w(z)$
crosses the ``phantom divide'' $w(z)=-1$ at $z\simeq0.464$.
The average value of $w(z)\simeq-1$ on the range $z\lsim0.7$, while the
average value of $w(z)<-1$ if the range of redshifts is extended to higher
values. This agrees with the evidence of the SneIa data.

In fact,
in a recent study \cite{ZZ} which examines constraints on the equation of
state by combining the Constitution SneIa data \cite{Hicken} with WMAP5
\cite{wmap} and SDSS constraints, Zhao and Zhang find 95\% confidence
level evidence in favour of a model with $w(z)>-1\in(0.25,0.5)$,
$w(z)<-1\in(0.5,0.75)$, meaning that $w(z)$ must cross the phantom divide
in the range $0.25<w<0.75$. The fiducial model of Fig.~\ref{fig_wz}(a)
crosses the phantom divide almost in the centre of this redshift range.

%------------------------------------------------------------
\begin{figure}[htb]
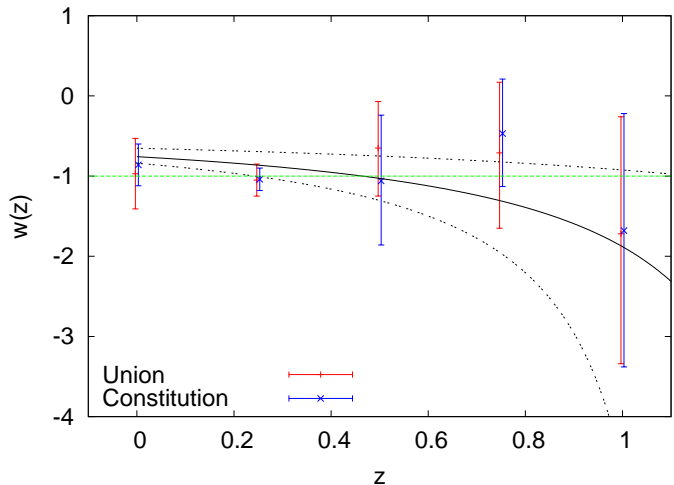

\vbox{\figwz
\caption{\label{fig:SCHMPS}%
{\sl The artificial equivalent of an equation of state (\ref{eosTS}) is
compared with a recent analysis of Serra \etal\ \cite{SCHMPS}. The third panel
of Fig.~1 of ref.\ \cite{SCHMPS} is combined with the curve of $w(z)$ for
the best--fit value $\fvn=0.76^{+0.12}_{-0.09}$ (solid curve; 1$\si$ limits
dotted curves). Following ref.\ \cite{SCHMPS} 2$\si$ uncertainties are plotted.
The $1\si$ uncertainties are tabulated in Table \ref{tab:SCHMPS}.}}}
\end{figure}
%------------------------------------------------------------
Another recent investigation \cite{SCHMPS} draws different conclusions
about evidence for dynamical dark energy. However, while the results of
Serra \etal\ \cite{SCHMPS} are consistent with a cosmological constant
at the 2$\si$ level, they are also consistent with the best--fit \name\ model
at the same level, as illustrated in Fig.~\ref{fig:SCHMPS}. The different
conclusions drawn by the authors of refs.\ \cite{ZZ} and \cite{SCHMPS}
result not only from using somewhat different data sets, but also from
differences in the treatments of data bins.

In considering Fig.~\ref{fig:SCHMPS} it should also be borne in mind
that there are significant systematic issues between the SALT and MLCS
data reduction methods, as will be discussed in Sec.~\ref{dis}. The Union
\cite{union} and Constitution \cite{Hicken} compilations use the SALT
method. A new analysis of 103 SDSS-II SneIa \cite{SDSSII} in the redshift
range, $0.04<z<0.42$ when combined with 185 SneIa from other surveys, yields
best--fit parameters $w=-0.96\pm0.06\w{(syst)}\pm{0.12}\w{(stat)}$ and
$\OmMn=0.265\pm0.016\w{(syst)}\pm{0.025}\w{(stat)}$ using SALT-II to fit to
spatially flat FLRW models with constant $w$, but
$w=-0.76\pm0.07\w{(syst)}\pm{0.11}\w{(stat)}$ and $\OmMn=0.307\pm
0.019\w{(syst)}\pm{0.023}\w{(stat)}$ using MLCS2k2. Use of MLCS data
reduction is therefore also likely to somewhat change the data values in
Table~\ref{tab:SCHMPS} and Fig.~\ref{fig:SCHMPS}.

At this stage the uncertainties, especially systematic ones in data
reduction, are too large to draw firm conclusions, but future measurements
may change the picture. Of course given specific models of dark energy,
greater statistical leverage is obtained simply by comparing
$\Hm D$ directly on a model by model basis.
\begin{table}
\begin{center}
 \begin{tabular}{|c|c|c|c|}
\hline
\ redshift &\ $w\Ns U$\ &\ $w\Ns C$\ &\ $w\Ns{TS}\vphantom{w\Z{\Z C}}$\ \\
\hline
0.0\ &\ $-0.97\pm0.22$&\ $-0.86\pm0.13$\ &\ $-0.76^{+0.11}_{-0.07}$\ \\
0.25 &\ $-1.05\pm0.10$&\ $-1.04\pm0.07$\ &\ $-0.86^{+0.17}_{-0.15}$\ \\
0.5\ &$-0.65^{+0.29}_{-0.30}$&$-1.06^{+0.41}_{-0.40}$&\ $-1.02\pm0.28$\ \\
0.75 &$-0.71^{+0.44}_{-0.47}$&$-0.47^{+0.34}_{-0.33}$&$-1.31^{+0.49}_{-0.65}$\\
1.0\ &$-1.72^{+0.73}_{-0.81}$&$-1.68^{+0.73}_{-0.85}$&$-1.88^{+0.96}_{-2.76}
\vphantom{w\Z{\Z C}}$\\
\hline
\end{tabular}
\caption{Values of $w(z)$ determined by Serra \etal\ \cite{SCHMPS} using a
standard FLRW cosmology are compared to the artificial equivalent of $w(z)$
for the \name\ model: $w\Ns U$ and $w\Ns C$ are the values determined by
combining WMAP5 CMB and SDSS-DR7 BAO data with the Union and Constitution
SneIa data sets respectively, as given in
Table I of ref.\ \cite{SCHMPS}. The equivalent $w\Ns{TS}$ for the \name\
model uses a void fraction $\fvn=0.76^{+0.12}_{-0.09}$ as determined
in ref.\ \cite{LNW}. 1$\si$ uncertainties are listed in each case.
\label{tab:SCHMPS}
}
\end{center}
\end{table}

\subsection{Angular--size redshift relation}

The angular size, $\de=\ell/\dA$, of a class of objects of uniform proper
length, $\ell$, is readily determined from (\ref{lumd}), (\ref{FF}).
Empirically the differences from the \LCDM\ model are not very large. For
the best-fit value $\fvn=0.762$ the minimum angle occurs at $z=1.74$, as
opposed to $z=1.67$ for a spatially flat \LCDM\ model with $\OmMn=0.249$,
or $z=1.56$ for a spatially flat \LCDM\ model with $\OmMn=0.34$ (see
Fig.~\ref{fig_angd}). The angle subtended by standard rulers in the \name\
model is very slightly less than for the comparison spatially flat \LCDM\
models. At $z=6$ the difference is of order 9--15\%.
%------------------------------------------------------------
\begin{figure}[htb]
\vbox{\figangd
\caption{\label{fig_angd}%
{\sl The angle, $\de$ (in arcsec), subtended by a 10kpc source as a function
of redshift for the \name\ model with $\fvn=0.762$, $\Hm=61.7\kmsMpc$ (solid
line) as compared to the equivalent angular size relation for three spatially
flat \LCDM\ models  (dashed lines from top to bottom): {\bf(a)}
$(\OmMn,\OmLn)=(0.279,0.721)$, $\Hm=71.9\kmsMpc$;
{\bf(b)} $(\OmMn,\OmLn) =(0.249,0.751)$, $\Hm=71.9\kmsMpc$;
{\bf(c)} $(\OmMn,\OmLn)=(0.34,0.66)$, $\Hm=62.7\kmsMpc$.}}}
\end{figure}
%------------------------------------------------------------

\section{The $H(z)$ and $Om(z)$ measures\label{Hom}}

Recently Sahni, Shafieloo and Starobinsky \cite{SSS} have proposed a new
diagnostic of dark energy \cite{foot4}, the function
\beq
Om(z)\equiv{{H^2(z)\over\Hm^2}-1\over(1+z)^3-1}\,,
\label{dSSS}\eeq
on account of the fact that it is equal to the constant present epoch
matter density parameter, $\OmMn$, at all redshifts for a spatially flat
FLRW model with pressureless dust and a cosmological constant, but is not
constant if the cosmological constant is replaced by other forms of dark
energy. For a spatially flat universe with pressureless dust plus some
arbitrary dark energy one has
\beq
Om(z)=\OmMn+(1-\OmMn){(1+z)^{3(1+w)}-1\over(1+z)^3-1}\,.
\eeq
For general FLRW models $H=D'^{-1}\sqrt{1+\Omkn\Hm^2 D^2}$ only
involves a single derivative of $D(z)$, and so the diagnostic (\ref{dSSS})
is easier to reconstruct observationally than the equation
of state parameter, $w(z)$.
%------------------------------------------------------------
\begin{figure}[htb]
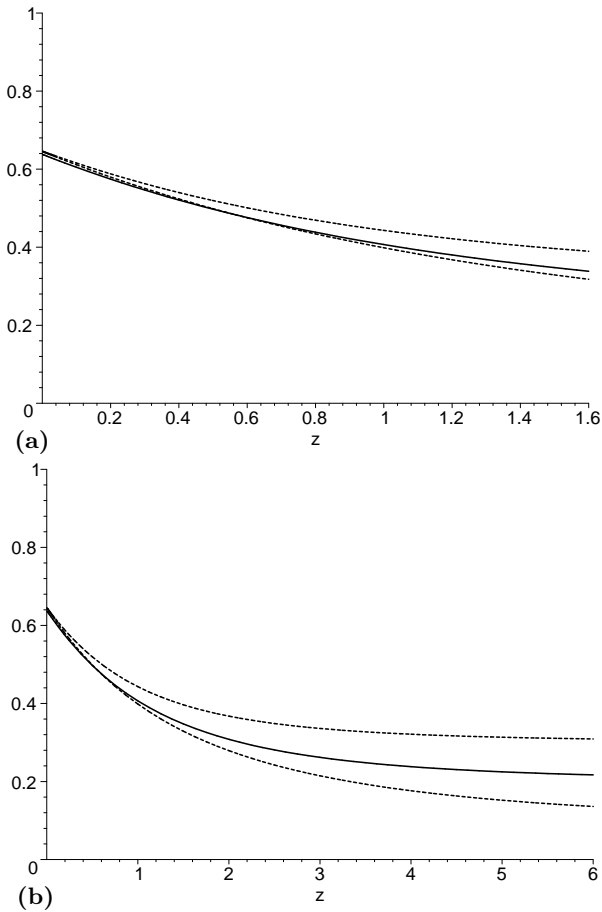

\vbox{\figOm
\caption{\label{fig_Om}%
{\sl The dark energy diagnostic $Om(z)$ of Sahni, Shafieloo and Starobinsky
\cite{SSS} plotted for the \name\ tracker solution with best--fit value
$\fvn=0.762$ (solid line), and $1\si$ limits (dashed lines) from ref.\
\cite{LNW}: {\bf(a)} for the redshift range $0<z<1.6$ as shown in
ref.\ \cite{SSS2}; {\bf(b)} for the redshift range $0<z<6$.}}}
\end{figure}
%------------------------------------------------------------

The quantity $Om(z)$ is readily calculated for the \name\ model, and is
shown in Fig.~\ref{fig_Om}. What is
striking about Fig.~\ref{fig_Om}, as compared to the curves for quintessence
and phantom dark energy models plotted in ref.\ \cite{SSS}, is that the
$z=0$ intercept
\beq
Om(0)=\frn23\left.H'\right|_0={2(8\fvn^3-3\fvn^2+4)(2+\fvn)\over(4\fvn^2+\fvn
+4)^2}
\label{Omint}\eeq
is substantially larger than in the dark energy models. We note from
(\ref{Omint}) that $\lim_{\fvn\to0}Om(0)=1$, and $\lim_{\fvn\to1}Om(0)=\frn23$,
with a minimum value of $Om(0)\simeq0.638$ at
$\fvn\simeq0.774$. The best-fit present epoch void fraction \cite{LNW}
gives a value of $Om(0)$ very close to the minimum. For the range
$\fvn=0.76^{+0.12}_{-0.9}$ \cite{LNW} $Om(0)$ is tightly constrained to
the range $0.638<Om(0)<0.646$.

A further difference for
the \name\ model is that $Om(z)$ does not asymptote to the dressed density
parameter $\OmMn$ in any redshift range. For quintessence models $Om(z)>\OmMn$,
while for phantom models $Om(z)<\OmMn$, and in both cases $Om(z)\to\OmMn$
as $z\to\infty$. In the \name\ model, $Om(z)>\OmMn\simeq0.33$ for $z\lsim1.7$,
while $Om(z)<\OmMn$ for $z\gsim1.7$. It thus behaves more like a quintessence
model for low $z$, in accordance with Fig.~\ref{fig_wz}. However, the
steeper slope and the completely different behaviour at large $z$ mean the
diagnostic is generally very different to that of dark energy models. For
large $z$,
\beq
\lim_{z\to\infty}Om(z)={2(1-\fvn)(2+\fvn)^3\over(4\fvn^2+\fvn+4)^2}\,,
\eeq
giving a value $\OMMn<Om(\infty)<\OmMn$, if $\fvn>0.25$. For example for
$\fvn=0.762$, we find $Om(\infty)\simeq0.2$.

Shafieloo, Sahni and Starobinsky \cite{SSS2} have recently tested the $Om(z)$
statistic against CMB, BAO and SneIa data, including the
Constitution SneIa data \cite{Hicken}. In comparing
their results with Fig.~\ref{fig_Om} it should be noted that their analysis
entails taking particular empirical functions for $w(z)$, and then
best--fitting the free parameters. The two functions they choose are:
{\bf(i)}\ $w(z)=w\Z0+w\Z1 z/(1+z)$; and {\bf(ii)}\
$w(z)=-\frn12[1+\tanh((z-z_t)\Delta)]$, where $w\Z0$, $w\Z1$, $z_t$ and
$\Delta$ are empirically fit constants. In both
these cases $w(z)$ is monotonic and cannot completely
accommodate the equivalent ``artificial dark energy
equation of state'' for the \name\ model as depicted in Fig.~\ref{fig_wz}
at large $z$. Furthermore, the effective $w(z)$ of
Fig.~\ref{fig_wz}(a) becomes nonlinear 
in the range $0.5\lsim z\lsim1$, contradicting the parameterization
of case (i) of \cite{SSS2}. Also, it crosses the ``phantom divide''
at $z\simeq0.464$ contradicting the parameterization of case (ii).
However, we can expect that the empirical
forms of $w(z)$ assumed by Shafieloo \etal\ to have some
comparative value for the \name\ model at very low values of $z$.
The greatest leverage should
come as $z\to0$. It is therefore interesting to note that of the two
empirical forms for $w(z)$ assumed by Shafieloo \etal, the one that
provides the better fit, case (ii), gives a best-fit intercept $Om(0)$
remarkably close to the expectation from (\ref{Omint}) for $\fvn=0.762$,
viz.\ $Om(0)=0.638$. (See Fig.~3 right hand panel of ref.\ \cite{SSS2}.)
Since this is not the expectation for either
a typical quintessence or phantom energy model, it is an encouraging
result for the \name\ model, which is also consistent with the
study of Zhao and Zhang \cite{ZZ}.

Shafieloo \etal\ have suggested \cite{SSS2} that their analysis of the
recent data might give a hint that ``dark energy is decaying''.
Given that the results of ref.\ \cite{SSS2} appear to
be consistent with the expectations of the \name\ model, our
analysis sheds a different light on this interpretation. It should also
be noted that a pure FLRW model with substantial
negative spatial curvature, i.e., with $\Omkn>0$, will give an intercept
$Om(0)=\OmMn+\frn23\Omkn$, whose value could assume a similar value to
that obtained for the \name\ model. Of course, this would require
a value of $\Omkn$ which is ruled out by the WMAP analysis for the FLRW
case, which is why such values have not been considered by Shafieloo
\etal. As observed above $Om(0)$ has a very tight range of values for
a wide range of reasonable values of $\fvn$. Thus if the tests of
the $Om(z)$ statistic could be improved to include a wider range of
empirical $w(z)$ functions, including those that more closely mimic
our relation (\ref{eosTS}), then this would be an interesting test
once significantly more data becomes available.

%------------------------------------------------------------
\begin{figure}[htb]
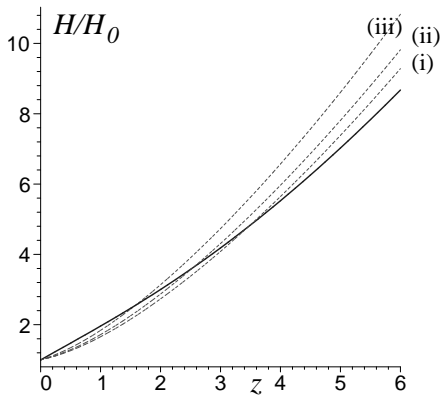

\vbox{\figHH0
\caption{\label{fig_HH0}%
{\sl The function $\Hm^{-1} H(z)$ for the \name\ model with
$\fvn=0.762$ (solid line) is compared to $\Hm^{-1}H(z)$ for three 
spatially flat \LCDM\ models with the same values of $(\OmMn,\OmLn)$ as in
Fig.~\ref{fig_coD} (dashed lines): {\bf(i)} $(\OmMn,\OmLn) =(0.249,0.751)$;
{\bf(ii)} $(\OmMn,\OmLn)=(0.279,0.721)$;
{\bf(iii)} $(\OmMn,\OmLn)=(0.34,0.66)$.}}}
\end{figure}
%------------------------------------------------------------
The strong differences seen in the $Om(z)$ diagnostic between
the \name\ model and typical dark energy models might be seen to
arise from the fact that it accentuates the differences which already
exist in the dressed $H(z)$ function, which is quite different to that
of the Friedmann equation. Using (\ref{HbH}) we plot $H(z)/\Hm$ for
the best--fit \name\ model, and compare it to the spatially flat \LCDM\
models that were plotted in Fig.~\ref{fig_coD}. For $z<1.5$ $H(z)/\Hm$
for the \name\ model with $\fvn=0.762$ is greater than for the \LCDM\
models shown. The absolute value of $H(z)$ is partly compensated for,
however, by the higher value of $\Hm$ that is generally assumed for the
\LCDM\ models.

Gazta\~naga, Cabr\'e and Hui \cite{GCH} have recently given measurements
of $H(z)$ at three redshifts, inferred from the separation of radial and
transverse BAO scales in the SDSS-DR6 data, as will be discussed in
Sec.\ \ref{APBAO}. However, their values are model dependent, being
estimated according to $$H(z)\ns{true}={r\Ns{BAO}\over r\Ns{WMAP}}
\Hm\sqrt{0.25(1+z)^3+0.75}$$
with a fiducial expansion rate for a spatially flat \LCDM\ model, with
$\OmMn=0.25$, used to convert redshifts to distances. Any estimates of
$H(z)$ will inevitably involve some model dependence, unless one can perform
a test such as the time drift of cosmological redshifts, which will be
discussed in Sec.\ \ref{drift}.

\section{The Alcock--Paczy\'nski test and baryon acoustic oscillations%
\label{APBAO}}

Alcock and Paczy\'nski devised a test \cite{AP} which relies on comparing the
radial and transverse proper length scales of spherical standard volumes
comoving with the Hubble flow \cite{footH}. This test was originally
conceived to distinguish FLRW models with a cosmological constant from those
without a $\Lambda$ term. The test is free from many evolutionary effects,
but relies on one being able to remove systematic distortions due to
peculiar velocities.

For the \name\ model the Alcock--Paczy\'nski test function determined
from the dressed geometry is
\beq
f\Ns{AP}={1\over z}\left|\Deriv\de z\th\right|={HD\over z}
={3\tbB(1+z)\dA\over t\tbA^2z}\,,
\label{fAP}\eeq
where $t$ is given implicitly in terms of $z$ by (\ref{redshift}).
%------------------------------------------------------------
\begin{figure}[htb]
\vbox{\figAP
\caption{\label{fig_AP}%
{\sl The Alcock--Paczy\'nski test function $f\Ns{AP}={1\over z}\left|\Deriv
\de z\th\right|$ for the \name\ model with
$\fvn=0.762$ (solid line) is compared to $f\Ns{AP}$ for three 
spatially flat \LCDM\ models with the same values of $(\OmMn,\OmLn)$ as in
Fig.~\ref{fig_coD} (dashed lines): {\bf(i)}$(\OmMn,\OmLn) =(0.249,0.751)$;
{\bf(ii)} $(\OmMn,\OmLn)=(0.279,0.721)$;
{\bf(iii)} $(\OmMn,\OmLn)=(0.34,0.66)$. Two redshift ranges are shown:
{\bf(a)} $0<z<1$; {\bf(b)} $0<z<6$.}}}
\end{figure}
%------------------------------------------------------------

In Fig.~\ref{fig_AP} the Alcock--Paczy\'nski test function (\ref{fAP}) is
compared to that of spatially flat \LCDM\ model with different values of
($\OmMn$,$\OmLn$). The curve for the \name\ model has a distinctly
different shape to those of the \LCDM\ models, being convex. However,
the extent to which the curves can be reliably distinguished would
require detailed analysis based on the precision attainable with any
particular experiment.

Current detections of the BAO scale in clustering statistics of LRGs
\cite{bao}--\cite{Percival2}
can in fact be viewed as a variant of the Alcock--Paczy\'nski test,
as they make use of both the transverse and radial dilations of the
fiducial comoving BAO scale to present a measure
\beq
D\Z V=\left[zD^2\over H(z)\right]^{1/3}=Df\Ns{AP}^{-1/3}.
\label{BAOr}\eeq
In Fig.~\ref{fig_Dv} the BAO radial test function (\ref{BAOr}) is
compared to that the same spatially flat \LCDM\ models plotted in
Fig.~\ref{fig_AP}, for the same redshift ranges.
%------------------------------------------------------------
\begin{figure}[htb]
\vbox{\figDv
\caption{\label{fig_Dv}%
{\sl The BAO radial test function $\Hm D\Z V=\Hm Df\Ns{AP}^{-1/3}$
for the \name\ model with
$\fvn=0.762$ (solid line) is compared to $\Hm D\Z V$ for three 
spatially flat \LCDM\ models with the same values of $(\OmMn,\OmLn)$ as in
Fig.~\ref{fig_coD} (dashed lines): {\bf(i)} $(\OmMn,\OmLn) =(0.249,0.751)$;
{\bf(ii)} $(\OmMn,\OmLn)=(0.279,0.721)$;
{\bf(iii)} $(\OmMn,\OmLn)=(0.34,0.66)$. Two redshift ranges are shown:
{\bf(a)} $0<z<1$; {\bf(b)} $0<z<6$.}}}
\end{figure}
%------------------------------------------------------------

Although the $D\Z V$ measure for the \name\ model is significantly different
to that of the \LCDM\ model at the higher redshifts shown in
Fig.~\ref{fig_Dv}(b), we see from Fig.~\ref{fig_Dv}(a) that at the nearby
redshifts the $D\Z V$ measure gives considerably less discriminatory
leverage. A case in point is provided by the ratio $\fdv\equiv
D\Z V(0.35)/D\Z V(0.2)$,
which has been determined observationally \cite{Percival,Percival2}. In this
case the \name\ model with $\fvn=0.76^{+0.12}_{-0.09}$ gives
$\fdv=1.632^{+0.005}_{-0.007}$, as compared to
$\fdv=1.664^{+0.009}_{-0.007}$ for a spatially flat
\LCDM\ model with $\OmMn=0.28\pm0.03$; values which are close. By comparison
the observed ratio was initially estimated to be $\fdv=1.812\pm0.060$
\cite{Percival}, but this estimate has recently been revised to
$\fdv=1.736\pm0.065$ \cite{Percival2}.

In fact, one must exercise caution in comparing the prediction of $\fdv$ for
the \name\ model with the ``observed'' ratio \cite{Percival,Percival2}
since the galaxy clustering data has been analysed in a manner which assumes
an underlying FLRW model.
The relevant analyses \cite{Percival,Percival2} involve transformations to
Fourier space to treat the power spectrum. To revisit such an analysis for
the \name\ model is far from trivial, as it requires a recalibration of
transfer functions, and of the cosmological drag epoch, $z_d$, for a rather
different cosmological parameterization given that we are dealing with
a model which does not evolve as a homogeneous isotropic cosmology. In
particular, the mass ratio of nonbaryonic dark matter to baryonic
matter can be somewhat different \cite{LNW} from the concordance \LCDM\ model,
and this needs to be   considered. Given that the difference in the value
of $\fdv$ quoted between refs.\ \cite{Percival} and \cite{Percival2} is due
to changes in the manner in which the data is treated, as well as the fact
that there is more data, it is clear the differences in calibration due to
a change of the nonbaryonic to baryonic mass ratio could also similarly
affect the value of the ``observed'' ratio.

A derivation of tools which would enable us to perform the BAO tests
to the extent of refs.\ \cite{Percival,Percival2} is beyond the scope of
the present paper. Instead, it is our aim to simply explore what the
best possible discriminating tests will be. In this regard, we note
that if we compare Fig.~\ref{fig_AP}(a) and Fig.~\ref{fig_Dv}(a), then
it is clear that the Alcock--Paczy\'nski test provides much more significant
differences between the \name\ model and \LCDM\ models than the $D\Z V$
measure. In fact, the $D\Z V$ measure is currently employed because there
is not yet sufficient data to separately estimate both the radial and
transverse BAO signals directly, as would be required for the
Alcock--Paczy\'nski test.

Gazta\~naga, Cabr\'e and Hui \cite{GCH} have recently
claimed to separate the radial and angular scales corresponding to the
BAO in the 2--point correlation function, by assuming a nonlinear
gravitational lensing magnification bias. Using SDSS-DR6 data they have
exhibited a correlation function in both the radial and transverse dimensions,
for redshift slices at $z=0.15$--$0.30$ and at $z=0.40$--$0.47$.
They have not yet provided separate estimates of both the radial and transverse
BAO scales. However, provided their techniques are robust, then a direct
Alcock--Paczy\'nski test may soon be on the horizon. Naturally such estimates
will have model dependence. From the the point of view of the \name\ model,
one must carefully consider not only the treatment of redshift space
distortions, but also any assumptions which rely on calibrations of FLRW
models, as discussed above.

\begin{table}
\begin{center}
 \begin{tabular}{|c|c|c|c|}
\hline
\ redshift &\ $\OmMn h^2$\ &\ $\OmBn h^2$\ &\ $\OmCn/\OmBn$\ \\
\ range &  &  &   \\
\hline
 0.15-0.30 & 0.132 & 0.028 & 3.7\\
 0.15-0.47 & 0.12\ & 0.026 & 3.6\\
 0.40-0.47 & 0.124 & 0.04\ & 2.1\\
\hline
\end{tabular}
\caption{Values of $\OmMn h^2$, $\OmBn h^2$ inferred by
Gazta\~naga, Cabr\'e and Hui \cite{GCH}, and the resulting mass ratio of
nonbaryonic dark matter to baryonic matter, $\OmCn/\OmBn$.
\label{tab:GCH}
}
\end{center}
\end{table}
One point of the analysis of Gazta\~naga, Cabr\'e and Hui is suggestive. 
They find some tension between their best--fit value of the baryon density
parameter $\OmBn\simeq0.06$ and the WMAP5 value \cite{wmap}
$\OmBn\simeq0.0432$. The discrepancy is greater in the higher redshift
slice. Their results are summarized in Table~\ref{tab:GCH}. The inferred
values for the mass ratio of nonbaryonic dark matter to baryonic
matter of $\OmCn/\OmBn=(\OmMn-\OmBn)/\OmBn$ are $3.6$ in the whole sample, 3.7
in the lower $z$ slice and 2.1 in the higher $z$ slice, as compared to
the expectation of a ratio of 6.1 from WMAP5, for which $(\OmBn h^2,
\OmMn h^2)=(0.0227,0.1308)$ \cite{wmap}. In other words, the best--fit
values indicate a somewhat higher mass fraction of baryons than the fit to
WMAP5 with a FLRW model. This is confirmed by analysis of the 3--point
correlation function \cite{GCCCF}, and is a feature which the authors
find difficult to explain as a systematic error. The analysis of
the 3--point function yields a best fit \cite{GCCCF} $\OmMn=0.28\pm0.05$,
$\OmBn=0.079\pm0.025$.

For the \name\ model by comparison, analysis of the Riess07 gold data
\cite{LNW,foot3} yields dressed parameters $\OmMn=0.33^{+0.11}_{-0.16}$,
$\OmBn=0.080^{+0.021}_{-0.013}$, and a ratio $\OmCn/\OmBn=3.1^{+2.5}_
{-2.4}$ from supernovae alone. Demanding a fit of the angular diameter
distance of the sound horizon \cite{clocks} to within 4\% would reduce
these bounds to $\OmCn/\OmBn=3.1^{+1.8}_{-1.3}$
for the \name\ model. Thus the higher baryon density indicated by the
analysis of Gazta\~naga, Cabr\'e and Hui is consistent with the
expectations of the \name\ model.

Finally, we note that although the reality of the BAO measure is
accepted by most researchers \cite{bao}--\cite{Seo}, Sylos Labini
\etal\ \cite{Sylos} have questioned this. Although Sylos Labini \etal\
detect the BAO scale in the LRG sample, they point out that its amplitude
is less than the overall density variations of 8\% at large scales, and
furthermore the correlation function remains positive where the
\LCDM\ model predicts it should be negative. Sample uncertainties may
limit the strength of this conclusion \cite{Kazin}, however.

In my view, although the results of ref.\ \cite{Sylos} may potentially
indicate problems with a statistical analysis based on the expectations
of a FLRW cosmology, the BAO is a real feature which will survive
despite the observed inhomogeneities. The point is that given a
universe which was very close to homogeneous and isotropic at last
scattering, it can only evolve so far away from homogeneity in the
time available since that epoch. Thus there is every reason to expect
that statistical analyses of the type that are being performed can pick
up a feature in the two--point correlation function, even if there are
larger scale variations in density of order 8\%. The exact properties of the
statistical correlation functions within a framework such as the \name\
cosmology await a detailed analysis. The main difference is that the
density of the observable universe when measured on scales larger than
that of ``statistical homogeneity'' will retain some intrinsic variance,
and furthermore is not the time--evolution of the mean density of the
statistical ensemble at last scattering. This is
likely to have important consequences for the statistical analysis.

\section{Test of (in)homogeneity\label{inhom}}

Recently Clarkson, Bassett and Lu \cite{CBL} have constructed what they call
a ``test of the Copernican principle'' based on the observation that
for homogeneous, isotropic models which obey the Friedmann equation,
the present epoch curvature parameter, a constant, may be written as
\beq
\Omkn={[H(z)D'(z)]^2-1\over[\Hm D(z)]^2}\label{ctest1}
\eeq
for all $z$, irrespective of the dark energy model or any other model
parameters. Consequently, taking a further derivative, the quantity
\beq
\CC(z)\equiv1+H^2(DD''-D'^2)+HH'DD'\label{ctest2}
\eeq
must be zero for all redshifts for any FLRW geometry.

A deviation of $\CC(z)$ from zero, or of (\ref{ctest1}) from a constant
value, would therefore mean that the assumption of homogeneity is violated.
Clarkson, Bassett and Lu refer to this as a ``violation of the Copernican
principle''. Given the viewpoint outlined in ref.\ \cite{clocks},
simply associating FLRW models with the Copernican principle is too great
a restriction on its general philosophy. One should distinguish the
Copernican Principle, which is generally understood as the statement that
we do not occupy a privileged position in the universe, from the
Cosmological Principle that the universe is described by a spatially
homogeneous isotropic geometry.

In the presence of inhomogeneity
there can still be statistically average cells -- taken here to
be of size $100h^{-1}$ Mpc --
but with a variance of the geometry within such cells. As observers
in an average galaxy, our position is unremarkable from the
point of view of the Copernican principle. Nonetheless, the local geometry
in an average void can be markedly different from the geometry in an
average galaxy. Given that observers and the things
they observe are necessarily in bound structures, structure formation
provides a selection
effect in terms of our local geometry vis--\`a--vis the volume--average
geometry in a void. Given this improved understanding of the Copernican
principle, one should not call the test of Clarkson, Bassett
and Lu a test of the Copernican principle. It is simply a test of the
validity of the FLRW models.  

Since the \name\ model is inhomogeneous, it will certainly violate the
test of Clarkson, Bassett and Lu. If one can determine $H(z)$ in a model
independent way, then tests of relations (\ref{ctest1}) or (\ref{ctest2})
could not only rule on whether the FLRW model is violated, but also
test the \name\ model. Analytic expressions for $HD'$ and $HD''$ are
obtained by multiplying (\ref{ctest1}) and (\ref{ctest2}) by (\ref{Ht}).
Combining the results with (\ref{lumd}) and (\ref{dHz}) we find that
(\ref{ctest1}) becomes
\beq
\Omkn={\BB(z)\over\Hm^2(1+z)^2\dA^2}
\eeq
where
\beq
\BB={(2t-b)\dA\over\tbA^2}\left[2+{(2t-b)\dA\over\tbA^2}\right]
\eeq
while (\ref{ctest2}) becomes
\beq
\CC=-{(2t-b)\dA\over\tbA^2}-{3b(10t^2+11bt-2b^2)\dA^2\over t\tbA^4}\,.
\eeq
Once again, $\dA$ is given by (\ref{lumd}) and $t$ is
given implicitly in terms of $z$ via (\ref{redshift}).

We plot the functions $\BB(z)$ and $\CC(z)$ in Figs.\ \ref{fig_Bex} and
\ref{fig_Cex}. The function $\CC$ differs appreciably from the FLRW
value of zero. However, two derivatives are required to determine $\CC(z)$,
which is subject to greater uncertainties for actual data, given that
$D(z)$ is effectively what is measured. Thus it would be more feasible
to determine $\BB(z)=[HD']^2-1$, which involves a single derivative of
the observed curve. (It makes more sense to plot $\BB(z)$, rather than the
right hand side of eq.\ (\ref{ctest1}), which involves a division by zero
as $z\to0$.) In Fig.~\ref{fig_Bex} $\BB(z)$ for the \name\ model is
compared to the expectations for \LCDM\ models with a small amount of
spatial curvature, as compatible with WMAP. The form of $\BB(z)$ is
very different at small redshifts, which suggests that this will
be a useful observational test. Furthermore, since $\BB(z)$ has a maximum
value and also changes sign, its form for the \name\ model is very different
to that of any FLRW model. In the FLRW case $\BB(z)$ is always a monotonic
function whose sign is determined by that of $\Omkn$. At large $z$, or
equivalently at early times as $t\to0$, $\BB(z)\to0$ and $\CC(z)\to0$
for the \name\ model, consistent with the fact that it coincides with a
spatially flat Einstein--de Sitter universe at early times. Since $\CC(z)$
involves second derivatives, it goes to zero more slowly than $\BB(z)$:
for the best-fit solution of Figs.\ \ref{fig_Bex} and \ref{fig_Cex},
$\BB(1100)\simeq-0.0029$, while $C(1100)\simeq0.075$.

It is interesting to compare Fig.~\ref{fig_Cex} with the corresponding plot of
$\CC(z)$ for a LTB model with a large void recently given in Fig.\ 14 of ref.\
\cite{FLSC}. The magnitude of $\CC(z)$ is considerably larger for the
\name\ model. 
%------------------------------------------------------------
\begin{figure}[htb]
\vbox{\figBex
\caption{\label{fig_Bex}%
{\sl The homogeneity test function $\BB(z)=[HD']^2-1$ is plotted for the
\name\ tracker solution with best--fit value $\fvn=0.762$ (solid line), and
compared to the equivalent curves $\BB=\Omkn(\Hm D)^2$ for two different
\LCDM\ models with small curvature: {\bf(a)} $\OmMn=0.28$, $\OmLn=0.71$ and
$\Omkn=0.01$; {\bf(b)} $\OmMn=0.28$, $\OmLn=0.73$ and $\Omkn=-0.01$.
A spatially flat FLRW model would have $\BB(z)\equiv0$.}}}
\end{figure}
%------------------------------------------------------------
\begin{figure}[htb]
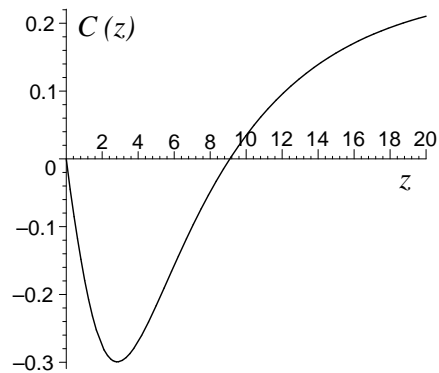

\vbox{\figCex
\caption{\label{fig_Cex}%
{\sl The homogeneity test function $\CC(z)$ given by (\ref{ctest2})
is plotted for the \name\ tracker solution with best-fit value $\fvn=0.762$.
Any FLRW model would have $\CC(z)\equiv0$, regardless of its spatial
curvature.}}}
\end{figure}
%------------------------------------------------------------

\section{Time drift of cosmological redshifts\label{drift}}

For the purpose of the (in)homogeneity test considered in the last section,
$H(z)$ must be observationally determined, and this is difficult to achieve
in a model independent way. There is one way of achieving this, however,
namely by measuring the time variation of the redshifts of different sources
over a sufficiently long time interval \cite{SML}, as has been discussed
recently in relation to tests of (in)homogeneity by Uzan, Clarkson and Ellis
\cite{UCE}. Although the measurement is extremely challenging, it may be
feasible over a 20 year period by precision measurements of the Lyman-$\al$
forest in the redshift range $2<z<5$ with the next generation of
Extremely Large Telescopes \cite{ELT}.

For FLRW models one has
\beq
\Deriv\dd t z=\Hm(1+z)-H(z)\label{dzdt1}
\eeq
which in the case of a \LCDM\ model with possible spatial curvature leads
directly to
\beq
{1\over\Hm}\Deriv\dd t z=(1+z)-\sqrt{\OmMn(1+z)^3+\OmLn+\Omkn(1+z)^2}.
\eeq
For the \name\ model one has an expression identical to (\ref{dzdt1})
in terms of the dressed Hubble parameter if the time derivative is
take with respect to wall time, $\ta$. Using (\ref{Ht}) we find that
\bea
{1\over\Hm}\Deriv\dd\ta z&=&1+z-{H\over\Hm}\nonumber\\
%&=&{2^{1/3}(4\fvn^2+\fvn+4)(t+b)\over\fvn^{1/3}(2+\fvn)\Hm t^{2/3}\tbA^{4/3}}
&=&1+z-{3\tbB\over\Hm t\tbA^2}\,,
\eea
where $t$ is given implicitly in terms of $z$ by (\ref{redshift}).
%------------------------------------------------------------
\begin{figure}[htb]
\vbox{\figzdot
\caption{\label{fig_zdot}%
{\sl The function $\Hm^{-1}\Deriv\dd\ta z$ for the \name\ model with
$\fvn=0.762$ (solid line) is compared to $\Hm^{-1}\Deriv\dd\ts z$ for three 
spatially flat \LCDM\ models with the same values of $(\OmMn,\OmLn)$ as in
Fig.~\ref{fig_coD} (dashed lines): {\bf(i)} $(\OmMn,\OmLn) =(0.249,0.751)$;
{\bf(ii)} $(\OmMn,\OmLn)=(0.279,0.721)$;
{\bf(iii)} $(\OmMn,\OmLn)=(0.34,0.66)$.}}}
\end{figure}
%------------------------------------------------------------

In Fig.~\ref{fig_zdot} we compare $\Hm^{-1}\Deriv\dd\ta z$ for the best-fit
\name\ model with $\fvn=0.762$ to the equivalent function for three different
spatially flat \LCDM\ models. What is notable is that the curve for
the \name\ model is considerably flatter than those of the \LCDM\ models.
The origin of this feature may be understood qualitatively to arise from
the fact that the magnitude of the apparent acceleration is considerably
smaller in the \name\ model, as compared to the magnitude of the acceleration
in \LCDM\ models. For models in which there is no apparent acceleration
whatsoever, one finds that $\Hm^{-1}\Deriv\dd\ta z$ is always negative.
If there is cosmic acceleration, real or apparent, at late epochs then
$\Hm^{-1}\Deriv\dd\ta z$ will become positive at low redshifts, though
at a somewhat larger redshift than that at which acceleration is deemed
to have begun.

Fig.~\ref{fig_zdot} demonstrates that a very clear signal of differences in
the redshift time drift between the \name\ model and \LCDM\ models might
be determined at low redshifts when $\Hm^{-1}\Deriv\dd\ta z$ should be
positive. In particular, the magnitude of $\Hm^{-1}\Deriv\dd\ta z$ is
considerably smaller for the \name\ model as compared to \LCDM\ models.
Observationally, however, it is expected that measurements
will be best determined for sources in the Lyman $\al$ forest in the range,
$2<z<5$. At such redshifts the magnitude of the drift is somewhat
more pronounced in the case of the \LCDM\ models. For a source at $z=4$,
over a period of $\de\ta=10$ years we would have $\de z=-3.3\times10^{-10}$
for the \name\ model with $\fvn=0.762$ and $\Hm=61.7\kmsMpc$. By comparison,
for a spatially flat \LCDM\ model with $\Hm=70.5\kmsMpc$ \cite{wmap} a source
at $z=4$ would over ten years give $\de z=-4.7\times10^{-10}$ for
$(\OmMn,\OmLn)=(0.249,0.751)$, and $\de z=-7.0\times10^{-10}$ for
$(\OmMn,\OmLn)=(0.279,0.721)$.

\section{Discussion\label{dis}}

In conclusion, the combination of tests we have described here
have the potential to decide between the \name\ cosmology, the
\LCDM\ cosmology, and other homogeneous isotropic cosmologies with
other sources of dark energy. A number of the tests have been devised
by other researchers with homogeneous dark energy cosmologies in mind.
In these cases, the results of independent analyses performed to
date are encouraging for the \name\ model. In particular,
\begin{itemize}
\item A study of $w(z)$ from recent datasets by Zhao and Zhang \cite{ZZ}
provides mild evidence at the 95\% confidence level for an effective $w(z)$
which crosses the ``phantom divide'' near the redshift $z\simeq0.46$
indicated in Fig.~\ref{fig_wz}(a), with $w(z)+1$ of the same sign over the
relevant redshift ranges for $z\lsim1$;
\item Fits of classes of empirical $w(z)$ functions by Shafieloo, Sahni and
Starobinsky \cite{SSS2} yield, in the best--fit case, an $Om(z)$ function with
intercept $Om(0)$ which appears to coincide with the \name\ expectation,
$Om(0)\simeq0.64$;
\item Studies of the BAO scale in SSDS-DR6 data by Gazta\~naga \etal\
\cite{GCH,GCCCF} yield a relative mass fraction of baryonic matter to
nonbaryonic dark matter, which is higher than the WMAP5 expectation with a
FLRW cosmology, but which is perfectly consistent with the \name\ model fit
to the angular scale of the sound horizon \cite{LNW}. 
\end{itemize}
While one can conceive of dark energy models with a $w(z)$ which mimics
Fig.~\ref{fig_wz}(a) at redshifts $z\lsim1$, there is no reason to expect
a different normalization of $\OmCn/\OmMn$ for such models. Indeed primordial
nucleosynthesis bounds are a very strong constraint on all cosmological
models. It is precisely because the mean CMB temperature at a volume--average
location unbound to physical structures in a void is cooler in the \name\
scenario than the mean
temperature we measure in a galaxy, that a different normalization of the
primordial baryon--to--photon ratio relative to present epoch cosmological
parameters is obtained. This would not be true for any homogeneous isotropic
cosmology, regardless of the type of dark energy fluid.

Other future tests discussed in this paper also have definitive predictions.
The expectation for the (in)homogeneity test of Clarkson, Bassett and Lu
\cite{CBL}, yields a diagnostic $\BB(z)$ which is both distinctively
different from FLRW models with spatial curvature as shown in
Fig.~\ref{fig_Bex}, and from LTB models. The time--drift of cosmological
redshifts would be most definitively tested by monitoring as many
redshifts as possible in the range $1\lsim z\lsim2$. As shown in
Fig.~\ref{fig_zdot}, in this range $\Hm^{-1}\Deriv\dd\ta z$ should be
very close to zero, and only very marginally positive as compared to the
\LCDM\ expectation. In the redshift range, $2\lsim z\lsim5$, which is
expected to be the range most readily tested with the next generation
of extremely large telescopes, the function $\Hm^{-1}\Deriv\dd\ta z$
will have a flatter $z$--dependence for the \name\ model than comparable
\LCDM\ models, as is seen in Fig.~\ref{fig_zdot}. The redshift range
$2\lsim z\lsim8$ can also be tested by GRB Hubble diagrams, and initial
investigations are in progress \cite{Schaefer}.

This paper has considered tests on scales greater than that of statistical
homogeneity. There are many other such tests in addition to those which we
have discussed. A number of these involve the CMB, such as
the determination of the amplitude of the late--time integrated Sachs--Wolfe
effect. Such tests require first a computation of the detailed structure
of the CMB acoustic peaks, recalibrated to the \name\ cosmology. This is a
very complicated task, which is why it has not been attempted here. However,
it is an important goal for future work.

Below the scale of statistical homogeneity we expect to see apparent
variance in the Hubble parameter, with a peak value 17\% larger than
the dressed global average value, measured over the scale of the
dominant void fraction of $30h^{-1}$ Mpc. Since voids dominate by volume,
a spherically symmetric average out to a fixed redshift will yield
generally higher values until we average over volumes for which a
typical line of sight intersects as many walls and voids as the
global average. That is, the spherically averaged Hubble parameter
should decrease from a maximum at the $30h^{-1}$ Mpc scale to the global
average value at roughly the $100h^{-1}$ Mpc scale. This general pattern
is indeed borne out by the analysis of Li and Schwarz \cite{LS2}. Much more
detailed predictions of the expected variance could be made for the
\name\ model, by performing Monte Carlo simulations assuming a reasonable
distribution of voids and minivoids packed into $100h^{-1}$ Mpc spheres.
This is an important goal for future work, as it would give a
Hubble bubble feature with unique characteristics, providing a test
of a feature for which there is no counterpart in the standard cosmology.

The recent determination of $\Hm=74.2\pm3.6\kmsMpc$ by the SH0ES survey
\cite{shoes} does provide a challenge for the \name\ model. However, as we
have just noted, in the \name\ scenario spatial curvature gradients and
apparent variance in the Hubble flow below the scale of statistical homogeneity
introduce systematic issues which complicate the determination of $\Hm$.
Riess \etal\ \cite{shoes} have done a very careful analysis, and make
efforts to account for a Hubble bubble -- which they cut off at
$z=0.023$, approximately two thirds of the scale of statistical
homogeneity. However, while they do not use supernovae with $z<0.023$
in the measurement of the Hubble flow, their calibration of the distance
ladder is necessarily made on nearby scales, in particular using the
maser distance to NGC 4258, at $7.2\pm0.5$ Mpc, as an anchor. In the \name\
scenario the effects of spatial inhomogeneity and spatial curvature gradients
are greatest on scales up to $30h^{-1}$ Mpc. Given that our own galaxy
appears to be in a filament, this may have an impact in calibrating standard
candles in the distance ladder.

The megamaser project \cite{mega} will therefore provide an interesting
test, as it will yield purely geometric distances -- independent
of standard candle calibrations -- on scales much larger than has been
tested to date \cite{footM}. The relevant scales are
considered to be well into the Hubble flow in the standard cosmology,
and if distances of order $\goesas60/h$ Mpc could be measured, would
represent a substantial up to a large fraction of the scale of statistical
homogeneity. The expectation in the \name\ scenario is that provided such
sources are sampled in directions in which the line of sight passes though a
variety of different density fields, then there should be variance in
the values of the Hubble constant so derived. The sample of maser
distances required to test the statistical expectations of the \name\
scenario would be considerably larger than the ten or so masers
currently under investigation, but may become feasible in coming decades.

In comparing future measurements with model predictions it is important
not only to extend the \name\ model to develop counterparts of all the
standard tests of the FLRW models, but also to carefully examine the methods
by which astronomical data is reduced, as in many cases the standard
cosmology is either explicitly or implicitly assumed. As one case in
point, BAO analyses at present typically use a transformation to Fourier
space and the use of spectral transfer functions calibrated to the
FLRW models. Thus while the results of Gazta\~naga \etal\ \cite{GCH,GCCCF}
are suggestive in that they find results in agreement with our
expected $\OmCn/\OmBn$ -- which is the physical parameter responsible for
the degree of baryon drag in the primordial plasma -- in applying results
of independent analyses to the \name\ model one must exercise caution until
each step in the BAO data reduction is understood directly from calibrations
with the \name\ model.

Another important case in which data reduction must be carefully
considered is that of supernovae. It was recently pointed out
\cite{KFL} that on Bayesian evidence the \name\ model is disfavoured as
compared to the \LCDM\ model using the Union \cite{union} and
Constitution \cite{Hicken} compilations. However, the Union and
Constitution data sets have been reduced using the SALT method in
which one simultaneously marginalizes over both empirical light curve
parameters and cosmological parameters, assuming a FLRW cosmology.
Hicken \etal\ \cite{Hicken} discuss and compare four different methods of
data reduction: SALT, SALT2, MLCS31 and MLCS17. They find some systematic
differences between the methods; for example, the SALT methods give
larger scatter at higher redshifts.

As will be discussed in a forthcoming
paper \cite{SW}, use of the MLCS17 reduced data gives a different
picture to the conclusions drawn by Kwan, Francis and Lewis \cite{KFL}.
In particular, analysis of the MLCS17--reduced 372 SneIa of Hicken
\etal\ gives Bayesian evidence which favours the \name\ model over the
\LCDM\ model. Thus there are already enough supernovae in principle to
distinguish between the models; except that systematic uncertainties
in the empirical methods by which standard candles are standardized at
present limit the conclusions that can be
drawn. Such issues are likely to also be a feature of many other
astrophysical observations, and thus it is important that as many
independent tests as possible are devised, and carried out carefully
in a way in which any model--dependent assumptions are scrutinized.

It is hoped that the tests discussed in this paper will provide a
basis for comparing the \LCDM\ model with a physically well--grounded
competing cosmological model. To fully compete, much further
development of the \name\ model is of course required. The standard
cosmology consists of a base model for expansion of the universe
-- the FLRW model dating from the 1920s -- on top of which a sophisticated
superstructure has been built over the last few decades. This superstructure
includes features such as the generation of initial conditions from inflation,
the bottom-up hierarchical structure formation process, and the results of
large-scale structure simulations using Newtonian gravity on top of the base
expansion. The model of refs.\ \cite{clocks,sol} replaces the base expansion
of the FLRW model by an average expansion which is not based on the Friedmann
equation, and this paper has explored a number of tests which can be
performed based solely on the average geometrical properties.

Many current cosmological tests of the standard cosmology -- including
detailed analysis of the CMB, galaxy clustering, redshift space
distortions and weak lensing -- can only be extended to the \name\
model once the standard cosmology superstructure built on top of the
FLRW model is adapted to the \name\ model to understand the growth
of structure at a more detailed level. Although this may seem a
daunting task, it is perhaps not as quite a tall order as one might
at first think. In particular, the differences from a standard FLRW
model with inflationary initial conditions at last scattering are
negligible, and consequently many large portions of the standard
cosmology would not change. In particular, the mechanisms of physical
processes are largely still the same, but what does change is the
relationship of present average cosmological parameters to the initial
perturbations. Rederivation of the standard cosmology superstructure
may largely be an issue of recalibration. Where the calculations involve
transfer functions that relate initial perturbation spectra to their
time evolved distributions, such recalibrations may be quite nontrivial,
however. Thus a careful first principles re-examination is required.
This is left to future work.

\medskip {\bf Acknowledgements} I would like to thank Teppo Mattsson,
Ishwaree Neupane and Peter Smale for discussions, and Jim Braatz, Thomas
Buchert, Chris Clarkson, Francesco Sylos Labini and Brad Schaefer for
correspondence. I also warmly thank Prof.\ Remo Ruffini and ICRANet for
support and hospitality while the bulk of the paper was completed. This
work was also partly supported by the Marsden fund of the Royal Society
of New Zealand.
\appendix
\section{General two--scale solution to the Buchert equations\label{apps}}

The general solution for the two--scale \cite{foot11} Buchert equations
(\ref{Beq1}), (\ref{Beq2}) for the independent functions $\ab(t)$ and
$\fv(t)$ is given implicitly by \cite{sol}
\beq
\fvf^{1/3}\ab=\ab\Z0\left[(1-\epi)\,\OMMn\right]^{1/3}
\left(\frac32\Hb t\right)^{2/3}\!,\label{sol1}
\eeq
\beq
\sqrt{u(u+\Ci)}-\Ci\ln\left(\left|u\over\Ci\right|^{1\over2}
+\left|1+{u\over\Ci}\right|^{1\over2}\right)
={\al\over\ab\Z0}\left(t+\te\right),\label{sol2}
\eeq
where
$u\equiv\fv^{1/3}\ab/\ab\Z0=\fvi^{1/3}\av/\ab\Z0$ is proportional to $\av$;
$\Ci\equiv\epi\OMMn\fvn^{1/3}/\OMkn$;
$\al\equiv\ab\Z0\Hb\OMkn^{1/2}/\fvn^{1/6}$;
$\epi$ and $\te$ being constants of integration, while $\fvn$,
$\Hb$, $\OMMn$ and $\OMkn$ are the present epoch values of $\fv$,
$\bH$, $\OMM$ and $\OMk$ respectively. Since $\fwi^{1/3}\aw=\fvf^{1/3}\ab$,
Eq.\ (\ref{sol1}) may also be written as $\aw=a\ns{w0}t^{2/3}$,
where $a\ns{w0}\equiv\ab\Z0\left[\frac94\fwi^{-1}(1-\epi)\,\OMMn{\Hb}^2
\right]^{1/3}$.

The lapse function, $\gb$, bare matter density parameter,
$\OMM$, and void fraction, $\fv$, satisfy the integral constraint
\beq{(1-\epi)\,\gb^2\OMM\over \fvf}=1.\label{Omtrue}\eeq
Furthermore, $\Hw=2/(3t)$, while $\Hv=\Hw/h_r$ where
\beq
h_r=\sqrt{(1-\epi)\OMMn\fvn^{1/3}\fv\over\left(\OMkn u+
\OMMn\fvn^{1/3}\epi\right)\fvf}\,.
\label{eqhr}\eeq

Of the six constants $\epi$, $\te$, $\fvn$, $\Hb$, $\OMMn$ and $\OMkn$,
only four are independent since there are additional constraints \cite{sol}
\beq
\sqrt{(1-\epi)\OMMn(1-\fvn)}+\sqrt{(\OMkn+\OMMn\epi)\fvn}=1\,
\label{con1}\eeq
\bea
&&{\OMkn^{3/2}\over\fvn^{1/2}}\Hb(\tn+\te)=\sqrt{\OMkn(\OMkn+\OMMn\epi)}
\nonumber\\ &&\ -\OMMn\epi\ln\left[\sqrt{\left|\OMkn\over\OMMn\epi
\right|}+\sqrt{\left|1+{\OMkn\over\OMMn\epi}\right|}\,\right]\!\!,
\label{con2}
\eea
where the age of the universe in volume--average time is
\beq
\tn={2\over3\Hb}\sqrt{1-\fvn\over(1-\epi)\OMMn}\,,
\eeq
on account of (\ref{sol1}).

Of the four independent parameters, two can be eliminated by demanding
priors at the surface of last scattering which are consistent with
the evidence of the CMB.
The redshift of the surface of last scattering relative to wall observers
at the present epoch, $z\simeq1100$, is fixed by the ratio of our locally
measured CMB temperature relative to the temperature scale of
matter--radiation decoupling and recombination, which is for the most part
determined by the binding energy of hydrogen. We require that the velocity
perturbations and density perturbations at this epoch when $z\simeq1100$
are consistent with observation. For example, we can fix velocity perturbations
by demanding $1-h_{ri}\simeq10^{-5}$ and density perturbations by
restricting $\fvi$. Physically, $\fvi$ is to be understood as the fraction
of our present horizon volume, $\cal H$, which by cosmic variance was in
uncompensated underdense perturbations at last scattering. If this
uncompensated fraction is viewed as a single density perturbation then
\beq
\de\Z{\cal H}\equiv\left(\de\rh\over\rh\right)_{\!\cal H\hbox{\sevenrm i}}
=\fvi\left(\de\rh\over\rh\right)\ns{vi}.\label{initc}\eeq
\smallskip

\noindent We might demand $\de\Z{\cal H}\in\{-10^{-6},-10^{-5}\}$, which
means we might take $\fvi\in\{10^{-4},10^{-2}\}$, depending on
what values of $(\de\rh/\rh)\ns{vi}$ are acceptable for the nonbaryonic dark
matter power spectrum.

Once values of $\hri$ and $\fvi$ are specified, then by (\ref{clocks2}),
(dropping the index $w$), $\gb_i=1-\fvi+\fvi\hri^{-1}$, while the initial
matter density parameter, $\Omi$, is fixed in terms of $\gb_i$, $\epi$ and
$\fvi$ by (\ref{Omtrue}). At the present epoch,
the integral constraint (\ref{Omtrue}), combined with the relation
for the cosmological redshift determined by wall observers, $z+1=\ab\Z0\gb
/(\ab\gc)$, gives
\beq
1-\fvn={(1-\epi)\OMMn\gb_i^2\fvi^{2/3}\OMkn^2\over(1+z_i)^2\fvn^{2/3}A_i^2}
\label{con3}\eeq
where $z_i\simeq1100$, and
\beq
A_i\equiv{\fvi^{1/3}\OMkn\ab_i\over\fvn^{1/3}\ab\Z0}=\OMMn\left[{\fvi(1-\epi)
\over(1-\fvi)\hri^2}-\epi\right]\,,
\eeq
where we have used (\ref{eqhr}) to express $\ab\Z0/\ab_i$ in terms of $\hri$
and other parameters in the last step.
We evaluate both (\ref{sol1}) and (\ref{sol2}) at the present epoch $\tn$
and at the time of last scattering, $t_i$, and compare them at each epoch
to eliminate $\tn$ and $\ts_i$. We then further eliminate $\te$ from the
two resulting expressions to also obtain
\begin{widetext}
\bea
&&\sqrt{A_i(A_i+\OMMn\epi)}-\OMMn\epi\ln\left(\sqrt{A_i\over\OMMn|\epi|}
+\sqrt{\left|1+{A_i\over\OMMn\epi}\right|}\right)
-{2\over3}\sqrt{(1-\fvi)A_i^3\over\fvn(1-\epi)\fvi\OMMn}\nonumber\\
=&&
\sqrt{\OMkn(\OMkn+\OMMn\epi)}-\OMMn\epi\ln\left(\sqrt{\OMkn\over\OMMn|\epi|}
+\sqrt{\left|1+{\OMkn\over\OMMn\epi}\right|}\right)
-{2\over3}\sqrt{(1-\fvn)\OMkn^3\over\fvn(1-\epi)\fvi\OMMn}
\label{con4}\eea
\end{widetext}
For fixed $z_i$, $\fvi$ and $\hri$ the combination of eqs.\ (\ref{con1}),
(\ref{con3}) and (\ref{con4}) determines three of the parameters $\{\OMMn,
\OMkn,\epi,\fvn\}$, leaving one independent parameter in addition to
$\Hb$. Of course, the values of $z_i$, $\fvi$ and $\hri$ which are consistent
with the observed CMB, do vary over some small ranges. Given the existence
of the tracker solution, however, these small variations do not significantly
affect macroscopic cosmological parameters. The macroscopic properties of the
universe depend significantly on the two independent parameters $\Hb$ and
$\fvn$.

\section{Tracker solution to the Buchert equations\label{appt}}

As noted in ref.\ \cite{sol}, setting $\epi=0$ in the general solution
gives a solution
which is a strong attractor in the phase space. Physically this solution
represents one in which the void regions expand as empty Milne universes
in volume average time, $\av=a\ns{v0}t$,
where $a\ns{v0}\equiv\OMkn^{1/2}\ab\Z0\Hb\fvn^{-1/6}\fvi^{-1/3}$, and
$h_r=2/3$. The solution is given by
\bea
\ab&=&{\ab\Z0\bigl(3\Hb t\bigr)^{2/3}\over2+\fvn}\left[3\fvn\Hb t+
(1-\fvn)(2+\fvn)\right]^{1/3}\nonumber\\
\label{track1}\\
\fv&=&{3\fvn\Hb t\over3\fvn\Hb t+(1-\fvn)(2+\fvn)}\,,
\label{track2}
\eea
with two independent parameters $\Hb$ and $\fvn$.

All other quantities of interest may be determined from (\ref{track1})
and (\ref{track2}). For example, the parameters (\ref{omm})--(\ref{omq})
are given by
\bea\OMM&=&{4\fvf\over(2+\fv)^2}={b(2t+3b)\over3(t+b)^2},\label{Omm}\\
\OMk&=&{9\fv\over(2+\fv)^2}={t(2t+3b)\over2(t+b)^2},\\
\OMQ&=&{-\fv\fvf\over(2+\fv)^2}={-bt\over6(t+b)^2}\label{Omq},
\eea
where $b=(1-\fvn)(2+\fvn)/[9\fvn\Hb]$, as in (\ref{bb}). From
(\ref{Omm})--(\ref{Omq}) we obtain equivalent expressions for their present
epoch values, $\OMMn$, $\OMkn$ and $\OM\Z{\QQ0}$, in terms of $\fvn$ or
$\tn=(2+\fvn)/(3\Hb)$. The bare Hubble parameter, lapse function and dressed
Hubble are given respectively by
\bea \bH&=&{2+\fv\over3t}={2(t+b)\over t(2t+3b)}\,,\label{bHt}\\
\gb&=&\half(2+\fv)={3(t+b)\over(2t+3b)}\,,\\ %\vphantom{\Biggl[}\,,\\
H&=&{4\fv^2+\fv+4\over 6t}={3(2t^2+3bt+2b^2)\over t(2t+3b)^2}\,.\label{Ht}
\eea
It also follows that
\beq
H={\left(4\fv^2+\fv+4\right)\bH\over2(2+\fv)}={3(2t^2+3bt+2b^2)\bH
\over 2(t+b)(2t+3b)}\,.
\label{HbH}
\eeq

For a number of the tests described in the paper, it is necessary to
perform derivative of observational quantities with the respect to the
redshift, $z$, as given by (\ref{redshift}). For this purpose, a useful
intermediate step is provided by
\beq
\Deriv\dd z{t^{1/3}}={-t^{1/3}(2t+3b)(t+b)\over3(z+1)\tbB}\,\label{dydz}
\eeq
which follows from (\ref{redshift}). Thus, for example, by (\ref{redshift}),
(\ref{Ht}) and (\ref{dydz}),
\bea
\Deriv\dd z{H}&=&{6(t+b)(2t^3+3bt^2+6b^2t+3b^3)\over(z+1)t\tbB\tbA^2},
\label{dHz}
\\ &=&{3\fvn^{1/3}\Hb(2t^3+3bt^2+6b^2t+3b^3)\over(2t)^{1/3}\tbB\tbA^{2/3}}\,.
\nonumber\eea

The above expressions all involve volume--average time, $\ts$. To
relate them to wall time, $\tc$, which is assumed to be a good approximation
to the time measured by typical observers in galaxies, one has to invert
the relation
\beq
\tc=\frn23\ts+{4\OmMn\over27\fvn\Hb}\ln\left(1+{9\fvn\Hb t
\over4\OmMn}\right)\,, \label{tsol}
\eeq
where $\OmMn=\frn12(1-\fvn)(2+\fvn)$ is the present epoch dressed matter
density.

%-----------------------------------------------------------------

%-----------------------------------------------------------------

\begin{thebibliography}{66}
%--------- References ---------%

\bibitem{fit1}
G.F.R.~Ellis,
%{\em``Relativistic cosmology: Its nature, aims and problems''},
in B.~Bertotti, F.~de Felice and A.~Pascolini (eds), {\it General
Relativity and Gravitation}, (Reidel, Dordrecht, 1984) pp.~215--288.

\bibitem{fit2}
G.F.R.~Ellis and W.~Stoeger,
%{\em``The fitting problem in cosmology''},
\CQG{4} (1987) 1697.
%%CITATION = CQGRD,4,1697;%%

\bibitem{buch1}
T.~Buchert,
%{\em``On average properties of inhomogeneous fluids in general relativity. I:
%Dust cosmologies''},
\GRG{32}, 105 (2000). %[gr-qc/9906015].
%%CITATION = GRGVA,32,105;%%

\bibitem{Celerier}
M.N.~C\'el\'erier,
%{\em``Do we really see a cosmological constant in the supernovae data?''},
\AaA{353}, 63 (2000); %[arXiv:astro-ph/9907206].
%%CITATION = ASTRO-PH 9907206;%%
%{\em``The accelerated expansion of the Universe challenged by an effect of the
%inhomogeneities. A review''},
New Adv.\ Phys.\ {\bf1}, 29 (2007) [astro-ph/0702416].
%%CITATION = ASTRO-PH 0702416;%%

\bibitem{Tom1}
K.~Tomita,
%{\em``A local void and the accelerating universe''},
\MNRAS{326}, 287 (2001); %[arXiv:astro-ph/0011484].
%CITATION = ASTRO-PH 0011484;%%
%{\em``Analyses of type Ia supernova data in cosmological models with a local
%void''},
Prog.\ Theor.\ Phys.\ {\bf106}, 929 (2001). %arXiv:astro-ph/0104141.
%CITATION = ASTRO-PH 0104141;%%

\bibitem{Ras1}
S.~R\"as\"anen,
%{\em``Dark energy from backreaction''},
JCAP {02} (2004) 003. %[arXiv:astro-ph/0311257]
%%CITATION = ASTRO-PH 0311257;%%

\bibitem{kolb0}
E.W.~Kolb, S.~Matarrese, A.~Notari and A.~Riotto,
%{\em``The effect of inhomogeneities on the expansion rate of the universe''},
\PR{D 71}, 023524 (2005); %[arXiv:hep-ph/0409038]
%%CITATION = PHRVA,D71,023524;%%
E.W.~Kolb, S.~Matarrese and A.~Riotto,
%{\em``On cosmic acceleration without dark energy''},
New J.\ Phys.\ {\bf8}, 322 (2006). %[arXiv:astro-ph/0506534].
%%CITATION = ASTRO-PH 0506534;%%

\bibitem{Moffat}
J.W.~Moffat,
%{\em``Cosmic microwave background, accelerating universe and inhomogeneous
%cosmology''},J
JCAP {10} (2005) 012; %[astro-ph/0502110]
%%CITATION = ASTRO-PH 0502110;%%
%{\em``Late-time inhomogeneity and acceleration without dark energy''},
JCAP {05} (2006) 001. %[astro-ph/0505326]
%%CITATION = ASTRO-PH 0505326;%%

\bibitem{EB}
G.F.R.~Ellis and T.~Buchert,
%{\em``The universe seen at different scales''},
\PL{A 347}, 38 (2005). %gr-qc/0506106
%%CITATION = PHLTA,A347,38;%%

\bibitem{footK}
{The Buchert equations were developed by a number of researchers in various
steps. Buchert and Ehlers \cite{BE} first considered the averaging problem
in Newtonian cosmology, but also concluded that an equation equivalent to
(\ref{buche2}) remained true in full general relativity. An equation
equivalent to (\ref{buche1}) was independently discussed by Carfora and
Piotrkowska \cite{CP}. The set of equations (\ref{buche1})--(\ref{buche3})
were then discussed in the perturbative relativistic dust case by Russ, Soffel,
Kasai and B\"orner \cite{RSKB}. The properties of the full set of equations
(\ref{buche1})--(\ref{buche4}), in a fully non-perturbative relativistic
setting, were subsequently discussed by Buchert \cite{buch1} and also
extended to perfect fluids \cite{buch2}.}

\bibitem{BE}
T.~Buchert and J.~Ehlers,
%{\em``Averaging inhomogeneous Newtonian cosmologies''},
\AaA{320}, 1 (1997); %[astro-ph/9510056]
%%CITATION = AAEJA,320,1;%
J.~Ehlers and T.~Buchert,
%{\em``Newtonian cosmology in Lagrangian formulation: Foundations and
%perturbation theory''},
\GRG{29}, 733 (1997). %[astro-ph/9609036]
%%CITATION = GRGVA,29,733;%%

\bibitem{CP}
M.~Carfora and K.~Piotrkowska,
%{\em``A renormalization group approach to relativistic cosmology''},
\PR{D 52}, 4393 (1995). %gr-qc/9502021
%%CITATION = PHRVA,D52,4393;%%

\bibitem{RSKB}
H.~Russ, M.H.~Soffel, M.~Kasai and G.~B\"orner,
%``Age of the universe: Influence of the inhomogeneities on the global
%expansion-factor''},
\PR{D 56}, 2044 (1997). %astro-ph/9612218
%%CITATION = PHRVA,D56,2044;%%

\bibitem{buch2}
T.~Buchert,
%{\em``On average properties of inhomogeneous fluids in general relativity:
%Perfect fluid cosmologies''},
\GRG{33}, 1381 (2001). %[gr-qc/0102049]
%%CITATION = GRGVA,33,1381;%%

\bibitem{footZ}
{An alternative scheme to that of Buchert was proposed earlier by Zalaletdinov
\cite{Zal}, which applies to general macroscopic averages in general
relativity. It averages all of the Einstein equations, and so an additional
integrability condition is not required. However, to make contact with
physical situations, such as cosmology, many additional assumptions are
required \cite{CPZ}--\cite{PS}.}

\bibitem{Zal}
R.M.~Zalaletdinov,
%{\em``Averaging out the Einstein equations and macroscopic space-time
%geometry''},
\GRG{24}, 1015 (1992);
%%CITATION = GRGVA,24,1015;%%
%{\em``Towards a theory of macroscopic gravity''},
{\bf25}, 673 (1993).
%%CITATION = GRGVA,25,673;%%

\bibitem{CPZ}
A.A.~Coley, N.~Pelavas and R.M.~Zalaletdinov,
%{\em``Cosmological solutions in macroscopic gravity''},
\PRL{95}, 151102 (2005).
%%CITATION = PRLTA,95,151102;%%

\bibitem{vdH}
R.J.~van den Hoogen,
%``A complete cosmological solution to the averaged Einstein field equations
%as found in macroscopic gravity,''
J.\ Math.\ Phys.\ {\bf 50}, 082503 (2009). %[arXiv:0909.0070]
%%CITATION = JMAPA,50,082503.%%

\bibitem{PS}
A.~Paranjape and T.P.~Singh,
%``The spatial averaging limit of covariant macroscopic gravity - Scalar
%corrections to the cosmological equations,''
\PR{D 76}, 044006 (2007); %[gr-qc/0703106]
%%CITATION = PHRVA,D76,044006;%%
A.~Paranjape,
%{\em``Backreaction of cosmological perturbations in covariant macroscopic
%gravity,''
\PR{D 78}, 063522 (2008). %[arXiv:0806.2755]
%%CITATION = ARXIV:0806.2755;%%

\bibitem{footC}
{Recently many new approaches have been suggested to averaging
schemes -- see, e.g., \cite{Sus}--\cite{Kor}. Our focus here is not on the
details of the averaging scheme, but in the physical interpretation of average
cosmological parameters when the variance in local geometry is large.
Although there may be further modifications, it is hoped that since
Buchert's scheme is phenomeologically robust enough to encapsulate the
leading order changes to the average evolution from the growth of density
perturbations.}

\bibitem{Sus}
R.A.~Sussman,
%{\em``Quasi-local variables, non-linear perturbations and back-reaction in
%spherically symmetric spacetimes''},
arXiv:0809.3314.
%%CITATION = ARXIV:0809.3314;%%

\bibitem{Lar}
J.~Larena,
%{\em``Spatially averaged cosmology in an arbitrary coordinate system''},
\PR{D 79}, 084006 (2009). %arXiv:0902.3159
%%CITATION = PHRVA,D79,084006;%%

\bibitem{BBM}
I.A.~Brown, J.~Behrend and K.A.~Malik,
%{\em``Gauges and cosmological backreaction''},
arXiv:0903.3264.
%%CITATION = ARXIV:0903.3264;%%

\bibitem{CAL}
C.~Clarkson, K.~Ananda and J.~Larena,
%{\em``The influence of structure formation on the cosmic expansion''},
\PR{D 80}, 083525 (2009). %[arXiv:0907.3377]
%%CITATION = ARXIV:0907.3377;%%

\bibitem{Coley}
A.A.~Coley,
%{\em``Averaging in cosmological models using scalars''},
arXiv:0908.4281.
%%CITATION = ARXIV:0908.4281;%

\bibitem{Kor}
M.~Korzy\'nski,
%{\em``Covariant coarse-graining of inhomogeneous dust flow in general
%relativity''},
arXiv:0908.4593.
%%CITATION = ARXIV:0908.4593;%%

\bibitem{IW}
A.~Ishibashi and R.M.~Wald,
%{\em``Can the acceleration of our universe be explained by the effects of
%inhomogeneities?''},
\CQG{23} (2006) 235. %[arXiv:gr-qc/0509108]
%%CITATION = GR-QC 0509108;%%

\bibitem{clocks}
D.L.~Wiltshire,
%{\em``Cosmic clocks, cosmic variance and cosmic averages''},
New J.\ Phys.\ {\bf9}, 377 (2007). %[gr-qc/0702082]
%%CITATION = NJOPF,9,377;%%

\bibitem{sol}
D.L.~Wiltshire,
%{\em``Exact solution to the averaging problem in cosmology''},
\PRL{99}, 251101 (2007). %[arXiv:0709.0732]
%%CITATION = PRLTA,99,251101;%%

\bibitem{essay}
D.L.~Wiltshire,
%{\em``Gravitational energy and cosmic acceleration''},
Int.\ J.\ Mod.\ Phys.\ {\bf D 17}, 641 (2008). %[arXiv:0712.3982]
%%CITATION = IMPAE,D17,641;%%

\bibitem{morph}
T.~Buchert, J.~Larena and J.M.~Alimi,
%{\em``Correspondence between kinematical backreaction and scalar field
%cosmologies: The morphon field''},
\CQG{23}, 6379 (2006); %[gr-qc/0606020]
%%CITATION = CQGRD,23,6379;%%
J.~Larena, J.M.~Alimi, T.~Buchert, M.~Kunz and P.S.~Corasaniti,
%{\em``Testing backreaction effects with observations''},
\PR{D 79}, 083011 (2009). %[arXiv:0808.1161]
%%CITATION = ARXIV:0808.1161;%%

\bibitem{Rasanen}
S.~R\"as\"anen,
%{\em``Accelerated expansion from structure formation''},
JCAP {11} (2006) 003; %[astro-ph/0607626]
%%CITATION = JCAPA,0611,003;%%
%{\em``Evaluating backreaction with the peak model of structure formation''},
JCAP {04} (2008) 026. %[arXiv:0801.2692]
%%CITATION = JCAPA,0804,026;%%

\bibitem{LS1}
N.~Li and D.J.~Schwarz,
%{\em``On the onset of cosmological backreaction''},
\PR{D 76}, 083011 (2007). %[gr-qc/0702043]
%%CITATION = PHRVA,D76,083011.%%

\bibitem{LS2}
N.~Li and D.J.~Schwarz,
%{\em``Scale dependence of cosmological backreaction''},
\PR{D 78}, 083531 (2008). %[arXiv:0710.5073]
%%CITATION = ARXIV:0710.5073;%%

\bibitem{Mattsson}
T.~Mattsson,
%{\em``Dark energy as a mirage''},
Gen.\ Relativ.\ Grav., in press; arXiv:0711.4264.
%%CITATION = ARXIV:0711.4264;%%

\bibitem{RF}
E.~Rosenthal and E.E.~Flanagan,
%{\em``Cosmological backreaction and spatially averaged spatial curvature''},
arXiv:0809.2107.
%%CITATION = ARXIV:0809.2107;%%

\bibitem{equiv}
D.L.~Wiltshire,
%{\em``Cosmological principle of equivalence and the weak--field limit''},
\PR{D 78}, 084032 (2008). %arXiv:0809.1183.
%%CITATION = ARXIV:0809.1183;%%

\bibitem{LNW}
B.M.~Leith, S.C.C.~Ng and D.L.~Wiltshire,
%{\em``Gravitational energy as dark energy: Concordance of cosmological
%tests''},
\ApJ{672}, L91 (2008). %[arXiv:0709.2535]
%%CITATION = ARXIV:0709.2535;%%

\bibitem{Riess07}
A.G.~Riess \etal, %[Supernova Search Team Collaboration],
%{\it``New Hubble Space Telescope discoveries of type Ia supernovae at $z>1$:
%Narrowing constraints on the early behavior of dark energy''},
\ApJ{659}, 98 (2007). %[astro-ph/0611572]
%%CITATION = ASTRO-PH 0611572;%%

\bibitem{foot1} The detailed comparison with more recent SneIa data sets
\cite{Hicken,union} is discussed in a forthcoming paper \cite{SW}.

\bibitem{union}
M.~Kowalski \etal,
%{\em``Improved cosmological constraints from new, old, and combined
%supernova data sets''},
\ApJ{686}, 749 (2008). %arXiv:0804.4142
%%CITATION = ASJOA,686,749;%%

\bibitem{Hicken}
M.~Hicken \etal,
%{\em``Improved dark energy constraints from ~100 new CfA supernova type Ia 
%light curves''},
\ApJ{700}, 1097 (2009). %[arXiv:0901.4804]
%%CITATION = ASJOA,700,1097;%%

\bibitem{SW}
P.R.~Smale and D.L.~Wiltshire,
in preparation.

\bibitem{wmap}
E.~Komatsu {\it et al.},
%{\em``Five-Year Wilkinson Microwave Anisotropy Probe (WMAP)
%observations: Cosmological interpretation''},
Astrophys.\ J.\ Suppl.\ {\bf 180}, 330 (2009). %[arXiv:0803.0547]
%%CITATION = ARXIV:0803.0547;%%

\bibitem{bao}
D.J.~Eisenstein \etal,
%{\em``Detection of the baryon acoustic peak in the large-scale correlation
%function of SDSS luminous red galaxies''},
\ApJ{633} (2005) 560; % [arXiv:astro-ph/0501171]
%%CITATION = ASTRO-PH 0501171;%%
S.~Cole \etal,
%{\em``The 2dF Galaxy Redshift Survey: power-spectrum analysis of the final
%data set and cosmological implications''},
\MNRAS{362} (2005) 505. % [arXiv:astro-ph/0501174]
%%CITATION = ASTRO-PH 0501174;%%

\bibitem{Percival}
W.J.~Percival \etal,
%S.~Cole, J.~Daniel, D.J.~Eisenstein, R.C.~Nichol, J.A~Peacock,
%A.C.~Pope and A.S.~Szalay, 2007,
%{\em``Measuring the Baryon Acoustic Oscillation scale using the SDSS and
%2dFGRS''},
\MNRAS{381}, 1053 (2007). %[arXiv:0705.3323]
%%CITATION = ARXIV:0705.3323;%%

\bibitem{Cabre}
A.~Cabr\'e and E.~Gazta\~naga,
%``Clustering of luminous red galaxies I: large scale redshift space
%distortions''},
\MNRAS{393}, 1183 (2009). %[arXiv:0807.2460]
%%CITATION = ARXIV:0807.2460;%%

\bibitem{Martinez}
V.J.~Mart\'{\i}nez \etal,
%``Reliability of the detection of the baryon acoustic peak''},
\ApJ{696}, L93 (2009); Erratum ibid.\ {\bf703}, L184 (2009). %[arXiv:0812.2154]
%%CITATION = ASJOA,696,L93;%%

\bibitem{Sanchez}
A.G.~S\'anchez, M.~Crocce, A.~Cabr\'e, C.M.~Baugh and E.~Gazta\~naga,
%{\em``Cosmological parameter constraints from SDSS luminous red galaxies: a
%new treatment of large-scale clustering''},
Mon.\ Not.\ R.\ Astr.\ Soc., in press; arXiv:0901.2570.
%%CITATION = ARXIV:0901.2570;%%

\bibitem{Percival2}
W.J.~Percival \etal,
%{\em``Baryon acoustic oscillations in the Sloan Digital Sky Survey Data
%Release 7 galaxy sample''},
arXiv:0907.1660.
%%CITATION = ARXIV:0907.1660;%%

\bibitem{Kazin}
E.A.~Kazin \etal,
%{\em``The baryonic acoustic feature and large-scale clustering in the SDSS LRG
%sample''},
arXiv:0908.2598.
%%CITATION = ARXIV:0908.2598;%%

\bibitem{Seo}
H.J.~Seo \etal,
%{\em``High-precision predictions for the acoustic scale in the non-linear
%regime''},
arXiv:0910.5005.
%%CITATION = ARXIV:0910.5005;%%

\bibitem{dark07}
D.L.~Wiltshire,
%{\em``Dark energy without dark energy''},
in {\em Dark Matter in Astroparticle and Particle Physics: Proceedings of
the 6th International Heidelberg Conference}, eds H.V.~Klapdor--Kleingrothaus
and G.F. Lewis, (World Scientific, Singapore, 2008) pp.~565-596
[arXiv:0712.3984].
%%CITATION = ARXIV:0712.3984;%%

\bibitem{reason}
{As discussed in Sec.\ 10 of ref.\ \cite{clocks}, in a void--dominated
universe one might expect a nearly fractal galaxy distribution below the
scale of statistical homogeneity. However, this is a likely
outcome of the model, rather than an input, and is limited by scale.
Since no assumptions about fractals are made in the model, I have decided
to rename it to avoid potential confusion. The word ``timescape'' captures
the idea that it is the relative calibration of ideal clocks in a nonuniform
dynamically evolving geometry which is a distinguishing feature of the
present cosmology.}

\bibitem{sfk}
J.E.~Forero--Romero, Y.~Hoffman, S.~Gottl\"ober, A.~Klypin and G.~Yepes,
%{\em``A dynamical classification of the cosmic web''},
\MNRAS{396}, 1815 (2009). %[arXiv:0809.4135]
%%CITATION = ARXIV:0809.4135;%

\bibitem{HV}
F.~Hoyle and M.S.~Vogeley,
%{\em``Voids in the Point Source Catalogue Survey and the Updated Zwicky
%Catalog''},
\ApJ{566}, 641 (2002); %[astro-ph/0109357].
%%CITATION = ASTRO-PH 0109357;%%
% \bibitem{hv2} F.~Hoyle and M.S.~Vogeley,
%{\em``Voids in the 2dF Galaxy Redshift Survey''},
\ApJ{607}, 751 (2004). %[astro-ph/0312533].
%%CITATION = ASTRO-PH 0312533;%%

\bibitem{minivoids}
A.V.~Tikhonov and I.D.~Karachentsev,
%{\em``Minivoids in the local volume''},
\ApJ{653}, 969 (2006). %[astro-ph/0609109]
%%CITATION = ASTRO-PH 0609109;%%

\bibitem{footD}
{In the standard cosmology distances are specified assuming Euclidean
geometry on nearby scales. In the \name\ scenario there are strong
spatial curvature gradients within the scale of statistical homogeneity,
and thus the Euclidean distance measure is not a good one strictly
speaking. We can make contact with the usual distance scales, however,
if we assume that distances are calibrated with the dressed geometry
(\ref{dgeom}) beyond the scale of statistical homogeneity. Below the
scale of statistical homogeneity we will quote the conventional
distances used by astronomers. However, the actual calibration of
distance measures below this scale is a subtle and an interesting
question still to be resolved.}

\bibitem{statcos}
A.~Gabrielli, F.~Sylos Labini, M.~Joyce and L.~Pietronero,
{\em Statistical Physics for Cosmic Structures},
(Springer, Berlin, 2005).

\bibitem{footL}
{It is of course envisaged that such a formalism will eventually be developed
in terms of contrasts with respect to various average effective
cosmological backgrounds. The models are simply still at an early stage
of development.}

\bibitem{Hogg}
D.W.~Hogg, D.J.~Eisenstein, M.R.~Blanton, N.A.~Bahcall, J.~Brinkmann, J.E.~Gunn
and D.P.~Schneider,
%{\em``Cosmic homogeneity demonstrated with luminous red galaxies''},
\ApJ{624}, 54 (2005). %[astro-ph/0411197]
%%CITATION = ASJOA,624,54;%%

\bibitem{Sylos}
F.~Sylos Labini, N.L.~Vasilyev, Y.V.~Baryshev and M.~L\'opez-Corredoira,
%{\em``Absence of anti-correlations and of baryon acoustic oscillations in the
%galaxy correlation function from the Sloan Digital Sky Survey DR7''},
\AaA{505}, 981 (2009). %[arXiv:0903.0950]
%%CITATION = ARXIV:0903.0950;%%

\bibitem{foot2}
{In the notation of Buchert and Carfora \cite{BC3}: $H\Z{\cal M}= H\ns{w}$;
$H\Z{\cal E}= H\ns{v}$; $\la\Z{\cal M}=1-\fv$; $\QQ\Z{\cal M}=0$
or $\de^2H\Z{\cal M}=\frn13\ave{\si^2}\Z{\cal M}$; $\QQ\Z{\cal E}
=0$ or $\de^2H\Z{\cal E}=\frn13\ave{\si^2}\Z{\cal E}$; and
$\QQ\Z{\cal D}=6\la\Z{\cal M}(1-\la\Z{\cal M})(H\Z{\cal E}-H\Z{\cal M})^2$.}

\bibitem{BC3}
T.~Buchert and M.~Carfora,
%{\em``On the curvature of the present-day Universe''},
\CQG{25}, 195001 (2008). %[arXiv:0803.1401]
%%CITATION = ARXIV:0803.1401;%%

\bibitem{footT}
{The best--fit values for the fit to the Riess07 data set obtained in
ref.\ \cite{LNW} were fit using the full exact solution for
$\hri=0.99999$, $\fvi=10^{-4}$, and $z_i=1100$. The difference in best-fit
parameters values between the full solution and the tracker solution in
this case turns out to be 0.1\% in $\Hm$, 1\% in $\fvn$,
and 3\% in $\OmMn$. Thus the tracker solution is accurate at the precision
of current tests. If higher precision is required it will also become necessary
to also incorporate radiation species.}

\bibitem{footQ}
{The magnitude of $\OMQ$ is of similar order to the volume--average
variance of the expansion rate found by Clarkson, Ananda and Larena
\cite{CAL} in an independent study, using a somewhat different averaging
scheme. In comparing results one should be careful to note that Clarkson
\etal\ consider domain averages on spatial hypersurfaces, when determining
the ``variance in the Hubble rate''. Such spatial volume averages relate to
our bare cosmological quantities, rather than dressed parameters. In the
present scheme one finds a greater variance in the apparent Hubble flow,
once one takes into account the fact that we dress parameters using rods
and clocks calibrated to our local geometry whose spatial curvature differs
from the spatial volume average one.}

\bibitem{foot5}
{In the terminology of Kolb, Marra and Matarese \cite{KMM} (\ref{dgeom})
is a ``phenomenological background solution'', whereas (\ref{avgeom}) is
an ``averaged background solution''.}

\bibitem{KMM}
E.W.~Kolb, V.~Marra and S.~Matarrese,
%{\em``Cosmological background solutions and cosmological backreactions''},
Gen.\ Relativ.\ Grav., in press; arXiv:0901.4566.
%%CITATION = ARXIV:0901.4566;%%

\bibitem{GRB}
B.E.~Schaefer,
%{\em``The Huble diagram to redshift >6 from 69 gamma-ray bursts''},
\ApJ{660}, 16 (2007); % arXiv:astro-ph/0612285
%%CITATION = ASTRO-PH/0612285;%%
N.~Liang, W.~K.~Xiao, Y.~Liu and S.N.~Zhang,
%{\em``A cosmology independent calibration of gamma-ray burst luminosity
%relations and the Hubble diagram''},
\ApJ{685}, 354 (2008); %[arXiv:0802.4262]
%%CITATION = ARXIV:0802.4262;%%
L.~Amati, C.~Guidorzi, F.~Frontera, M.~Della Valle, F.~Finelli, R.~Landi
and E.~Montanari,
%`{\em`Measuring the cosmological parameters with the Ep,i-Eiso correlation of
%gamma-ray bursts''},
\MNRAS{391}, 577 (2008); %[arXiv:0805.0377]
%%CITATION = ARXIV:0805.0377;%%
R.~Tsutsui, T.~Nakamura, D.~Yonetoku, T.~Murakami, Y.~Kodama and K.~Takahashi,
%{\em``Cosmological constraints from calibrated Yonetoku and Amati relation
%implies fundamental plane of gamma-ray bursts''},
JCAP {08} (2009) 015. %[arXiv:0810.1870]
%%CITATION = ARXIV:0810.1870;%%

\bibitem{Schaefer}
B.E.~Schaefer,
in preparation.

\bibitem{footN} Similar discontinuous $w(z)$ functions are found in
\LCDM\ models with a small amount of spatial curvature, $|\Omkn|=0.02$,
if $w(z)$ is mistakenly reconstructed with the assumption of a spatially flat
cosmology \cite{BFMV}.

\bibitem{BFMV}
G.~Barenboim, E.~Fern\'andez-Mart\'{\i}nez, O.~Mena and L.~Verde,
%{\em``The dark side of curvature''},
arXiv:0910.0252.
%%CITATION = ARXIV:0910.0252;%%

\bibitem{ZZ}
G.B.~Zhao and X.~Zhang,
%{\em``Probing dark energy dynamics from current and future cosmological
%observations,''
arXiv:0908.1568.
%%CITATION = ARXIV:0908.1568;%%

\bibitem{SCHMPS}
P.~Serra, A.~Cooray, D.E.~Holz, A.~Melchiorri, S.~Pandolfi and D.~Sarkar,
%{\em``No evidence for dark energy dynamics from a global analysis of
%cosmological data''},
arXiv:0908.3186.
%%CITATION = ARXIV:0908.3186;%%

\bibitem{SDSSII}
R.~Kessler \etal,
%``First-year Sloan Digital Sky Survey-II (SDSS-II) supernova results: Hubble
%diagram and cosmological parameters''},
Astrophys.\ J.\ Suppl.\ {\bf185}, 32 (2009). %[arXiv:0908.4274]
%%CITATION = ARXIV:0908.4274;%%

\bibitem{SSS}
V.~Sahni, A.~Shafieloo and A.A.~Starobinsky,
%{\em``Two new diagnostics of dark energy''},
\PR{D 78}, 103502 (2008). %[arXiv:0807.3548]
%%CITATION = ARXIV:0807.3548;%%

\bibitem{foot4}
{Equivalent diagnostics were also given independently by Gu, Chen and Chen
\cite{GCC}, and by Zunckel and Clarkson \cite{ZC}.}

\bibitem{GCC}
J.A.~Gu, C.W.~Chen and P.~Chen,
%{\em``A new approach to constraining quintessence models by observations''},
New J.\ Phys.\ {\bf11}, 073029 (2009). %[arXiv:0803.4504]
%%CITATION = ARXIV:0803.4504;%%

\bibitem{ZC}
C.~Zunckel and C.~Clarkson,
\PRL{101}, 181301 (2008). %[0807.4304]
%{\em``Consistency tests for the cosmological constant''},
%%CITATION = PRLTA,101,181301;%%

\bibitem{SSS2}
A.~Shafieloo, V.~Sahni and A.A.~Starobinsky,
%{\em``Is cosmic acceleration slowing down?''},
\PR{D 80}, 101301 (2009). %[arXiv:0903.5141]
%%CITATION = ARXIV:0903.5141;%%

\bibitem{GCH}
E.~Gazta\~naga, A.~Cabr\'e and L.~Hui,
%``Clustering of luminous red galaxies IV: Baryon acoustic peak in the
%line-of-sight direction and a direct measurement of H(z)''},
\MNRAS{399}, 1663 (2009). %[arXiv:0807.3551]
%%CITATION = ARXIV:0807.3551;%%

\bibitem{AP}
C.~Alcock and B.~Paczy\'nski,
%{\em``An evolution free test for non-zero cosmological constant''},
Nature {\bf281}, 358 (1979).
%%CITATION = NATUA,281,358;%%

\bibitem{footH}{A new variant of the Alcock--Paczy\'nski diagnostic has
been recently discussed by Garc\'{\i}a-Bellido and T.~Haugb{\o}elle
\cite{GBH} in the context of LTB models.}

\bibitem{GBH}
J.~Garc\'{\i}a-Bellido and T.~Haugb{\o}elle,
%{\em``The radial BAO scale and cosmic shear, a new observable for
%inhomogeneous cosmologies''},
JCAP {09} (2009) 028. %[arXiv:0810.4939]
%%CITATION = ARXIV:0810.4939;%%

\bibitem{GCCCF}
E.~Gazta\~naga, A.~Cabr\'e, F.~Castander, M.~Crocce and P.~Fosalba,
%{\em``Clustering of Luminous Red Galaxies III: Detection of the baryon
%acoustic  peak in the 3-point correlation function''},
\MNRAS{399}, 801 (2009). %[arXiv:0807.2448
%%CITATION = ARXIV:0807.2448;%%

\bibitem{foot3} The value of $\OmBn$ has been determined assuming a
tight range of values for the baryon--to--photon ratio at decoupling,
$\eta\Z{B\ga}=4.6$--$5.6\times10^{-10}$, as would be consistent with
primordial lithium abundance observations. The relatively large uncertainties
here reflect the $1\si$ uncertainties in $\OmMn$ which is not tightly
constrained by SneIa. Allowing a wider range of values for $\eta\Z{B\ga}$
would further increase the uncertainty in $\OmBn$. The estimates of $\OmMn$
and $\OmBn$ are coupled, and this has been taken into account in determining
$\OmCn/\OmBn$.

\bibitem{CBL}
C.~Clarkson, B.~Bassett and T.C.~Lu,
%{\em``A general test of the Copernican Principle''},
\PRL{101}, 011301 (2008). %[arXiv:0712.3457]
%%CITATION = PRLTA,101,011301;%%

\bibitem{FLSC}
S.~February, J.~Larena, M.~Smith and C.~Clarkson,
%{\em``Rendering dark energy void''},
arXiv:0909.1479.
%%CITATION = ARXIV:0909.1479;%%

\bibitem{SML}
A.~Sandage,
%{\em``The change of redshift and apparent luminosity of galaxies due to
%the deceleration of selected expanding universes''},
\ApJ{136}, 319 (1962);
%%CITATION = ASJOA,136,319;%%
G.C.~McVittie,
%{\em``Addition''},
\ApJ{136}, 334 (1962);
%%CITATION = ASJOA,136,334;%%
A.~Loeb,
%{\em``Direct measurement of cosmological parameters from the cosmic
%deceleration of extragalactic objects''},
\ApJ{499}, L111 (1998).
%%CITATION = ASTRO-PH/9802122;%%

\bibitem{UCE}
J.P.~Uzan, C.~Clarkson and G.F.R.~Ellis,
%{\em``Time drift of cosmological redshifts as a test of the Copernican
%principle''},
\PRL{100}, 191303 (2008). %[arXiv:0801.0068]
%%CITATION = PRLTA,100,191303;%%

\bibitem{ELT}
P.S.~Corasaniti, D.~Huterer and A.~Melchiorri,
%{\em``Exploring the dark energy redshift desert with the Sandage-Loeb
%test''},
\PR{D 75}, 062001 (2007); %[astro-ph/0701433]
%%CITATION = PHRVA,D75,062001;%%
J.~Liske {\it et al.},
%{\em``Cosmic dynamics in the era of Extremely Large Telescopes''},
\MNRAS{386}, 1192 (2008). %[arXiv:0802.1532]
%%CITATION = ARXIV:0802.1532;%%

\bibitem{shoes}
A.G.~Riess {\it et al.},
%``A redetermination of the Hubble constant with the Hubble Space Telescope
%from a differential distance ladder,''
\ApJ{699}, 539 (2009). %[arXiv:0905.0695]
%%CITATION = ASJOA,699,539;%%

\bibitem{mega}
M.J.~Reid, J.A.~Braatz, J.J.~Condon, L.J.~Greenhill, C.~Henkel and K.Y.~Lo,
%{\em``The Megamaser Cosmology Project: I. VLBI observations of UGC 3789''},
\ApJ{695}, 287 (2009). % [arXiv:0811.4345]
%%CITATION = ASJOA,695,287;%%

\bibitem{footM}
{The first distance of $49.5\pm7.5$ Mpc, announced at
http://www.nrao.edu/pr/2009/megamaser/, to galaxy UGC 3789 with a
recession velocity of $3325\kms$ \cite{mega}, would represents a value
$\Hm\simeq67.2\pm11.4\kmsMpc$, though this would change with Hubble flow
modelling (J.A.~Braatz, private communication). UGC 3789 happens to be
close to the distance scale at which we expect a peak in the average $\Hm$,
but also the maximum variance.}

\bibitem{KFL}
J.~Kwan, M.J.~Francis and G.F.~Lewis,
%{\em``Fractal bubble cosmology: A concordant cosmological model?''},
\MNRAS{399}, L6 (2009). %[arXiv:0902.4249]
%%CITATION = ARXIV:0902.4249;%%

\bibitem{foot11}
{Buchert and Carfora \cite{BC3} refer
to the same approximation as a ``three--scale model'', as they count the scale
of statistical homogeneity as one of the scales. My two scales refer to
largest relevant scales in the ``nonlinear regime''.}
%-----------------------------------------------------------------
\end{thebibliography}
\end{document}